\documentclass{aa} % for submitting

\usepackage{etex}
\usepackage{graphicx}
\usepackage{txfonts}

\usepackage{hyperref}
\hypersetup{
    colorlinks,
    linkcolor={blue!50!black},
    citecolor={blue},
    urlcolor={blue!50!black}
}
\usepackage{amsmath}   % Advanced maths commands
\usepackage{amssymb}   % Extra maths symbols
\usepackage{xcolor}     % colored text comments
\usepackage{floatrow}
\usepackage{booktabs}       % professional-quality tables

\providecommand{\LEt}[1]{}
\defcitealias{Boess2023b}{Paper~I}
\newcommand{\mrd}{\mathrm{d}}

\newcommand{\synchweb}{\citetalias{Boess2023b}}
\newcommand{\oblfig}{13}
\newcommand{\Ms}{\mathcal{M}_s}
\newlength{\fullwidth}
\newlength{\halfwidth}
\begin{document}

\title{Simulating the LOcal Web (SLOW)}

   \subtitle{VI. Gamma-ray emission in the local Universe}

   \author{Ludwig M. Böss \thanks{\email{lboess@uchicago.edu}}\fnmsep\inst{1,2}
          \and
          Ildar Khabibullin\inst{2,3}
          \and
          Daniel Karner\inst{2}
          \and
          Klaus Dolag\inst{2,3}
          \and \\
          Ulrich P. Steinwandel\inst{3}
          \and
          Elena Hernandez-Martinez\inst{2}
          \and
          Jenny G. Sorce\inst{4,5}
          }

   \institute{
Department of Astronomy and Astrophysics, The University of Chicago, William Eckhart Research Center, 5640 S. Ellis Ave. Chicago, IL 60637
\and
Universitäts-Sternwarte, Fakultät für Physik, Ludwig-Maximilians-Universität München, Scheinerstr.1, 81679 München, Germany
\and
Max Planck Institute for Astrophysics, Karl-Schwarzschild-Str. 1, D-85741 Garching, Germany
\and
%Center for Computational Astrophysics, Flatiron Institute, 162 5th Avenue, New York, NY 10010, USA
%\and
Univ. Lille, CNRS, Centrale Lille, UMR 9189 CRIStAL, F-59000 Lille, France
\and
Universit\'e Paris-Saclay, CNRS, Institut d'Astrophysique Spatiale, 91405, Orsay, France
             }

   \date{Received XXXX; accepted YYYY}

  \abstract
  {Diffuse $\gamma$-ray emission from cosmic-ray (CR) protons scattering off the gas in the intracluster and intergalactic medium remains out of reach for current observations. Detecting this emission would provide constraints on the nonthermal pressure support by CR protons in these environments.} %leave it empty if necessary  
   {We provide estimates for diffuse $\gamma$-ray emission in the \textit{Fermi}-LAT band from galaxy clusters and the cosmic web in the local Universe.}
   {In this work, we show results from the first cosmological magnetohydrodynamic simulation with an on-the-fly spectral CR model. We modeled CR injection at shocks, accounted for adiabatic energy changes and advection of CR protons, and obtained their $\gamma$-ray emissivity directly from the simulated CR energy density and spectra. 
   To do this, we used constrained initial conditions that evolved in a field closely resembling that of the local Universe, allowing a direct comparison to \textit{Fermi}-LAT data on massive clusters.}
   {We find CR proton acceleration at all structure formation and accretion shocks in galaxy clusters and cosmic web filaments. These protons provide the basis for diffuse $\gamma$-ray emission in these regimes. The absolute value of the diffuse $\gamma$-ray emission in our simulation lies a few orders of magnitude below the current upper limits found by \textit{Fermi}-LAT. Under the assumption of our model, a sensitivity of $F_\gamma < 10^{-11} \: \gamma~ \text{s}^{-1}~\text{cm}^{-2}$ would be required for a detection of diffuse emission in Coma. This provides a lower limit for diffuse emission from CR protons accelerated at structure formation shocks.}
   {}
   
\titlerunning{ Simulating the LOcal Web (SLOW) - VI}
\maketitle

\section{Introduction \label{sec:gammaweb_intro}}

Gamma-ray astronomy holds the potential to give insight into a number of nonthermal processes in the intracluster medium (ICM) and the intergalactic medium (IGM).
On the one hand, $\gamma$ photons emitted by blazars allow us to obtain a lower limit on the IGM magnetic field strength and correlation lengths (e.g. \citet{AlvesBatista2020, Acciari2022}, but see \citet{Broderick2018} for possible limitations of this method).
Direct observation of diffuse $\gamma$-ray emission in galaxy clusters, on the other hand, could provide insight into the relativistic proton component in the ICM (see \citealt{Bykov2019} for a recent review).
In galaxy clusters, the main mechanism of cosmic-ray proton (CRp)  production is diffusive shock acceleration (DSA) at shocks driven by mergers during the hierarchical structure formation process (see, e.g., \citealt{Brueggen2012, Brunetti2014} for reviews of galaxy cluster shocks).
These shocks have been studied in detail in the X-ray band through the shock-induced density and temperature increase \citep[e.g.,][for a review]{Boehringer2010}, and through radio emission attributed to efficient cosmic-ray (CR) electron acceleration in the so-called radio relics \citep[see][for a review]{Weeren2019}.
Because the energy-loss mechanisms of relativistic protons are inefficient \citep[see, e.g., Fig.~7 in][]{Brunetti2007}, they should be long-lived and should therefore be able to be transported across the cluster volume.

These relativistic protons can then interact with the thermal background gas and scatter into pions.
The neutral pion $\pi^0$ further decays into two $\gamma$ photons.
However, the current lack of a detection of such diffuse emission \citep[see, e.g.,][]{Ackermann2014, Ackermann2015, Ackermann2016} with the Fermi Large Area Telescope (\textit{Fermi}-LAT) places a limit of roughly 1\% of the thermal energy density on the relativistic proton-energy density \citep[however there are indications for diffuse $\gamma$-ray emission towards the Coma cluster by][]{Prokhorov2014, Xi2018, Adam2021, Baghmanyan2022}.

This places a strain on most theories for the efficiency of proton acceleration at merger shocks in the ICM \citep[see, e.g.,][]{Kang2007, Kang2013} that model CR acceleration efficiencies to match radio observations.
Under the assumption that CRp acceleration is more efficient than CR electron acceleration, this would produce CRps to thermal pressure ratios of up to 20\% \citep[see][for a recent review]{Wittor2021}.

More recent models include the shock obliquity, that is, the angle between shock propagation and background magnetic field, as it has been shown that different obliquities drive different types of plasma instabilities that favor either proton or electron acceleration \citep[e.g.,][]{Caprioli2014, Ha2018, Ryu2019}.
Because protons are found to be most efficiently accelerated at quasi-parallel shocks and because observations of radio relic polarization \citep[e.g.,][]{Stroe2013, Stroe2016, DiGennaro2018} and simulations \citep[e.g.,][]{Wittor2017, Banfi2020, Boess2023b} indicated that a large fraction of merger shocks are quasi-perpendicular, this would make merger shocks inefficient at accelerating CRps, which would help to alleviate this problem.

Previous studies have found that with an acceleration efficiency $\eta$ of up to 50\%, which is significantly above the currently favored efficiencies, the detection of diffuse $\gamma$-ray emission is just out of reach of \textit{Fermi}-LAT \citep[][]{Pinzke2010, Pinzke2011}.
More recent simulations by \citet{Vazza2016} showed that the acceleration efficiency should be on the order of $\eta \sim 10^{-3}$ to avoid disagreement with the non-detection by \textit{Fermi}-LAT.
This was followed up by \citet{Ha2020} with the most recent Mach-number-dependent DSA parameterization of \citet{Ryu2019}, which gave a range of $\eta \approx 10^{-3} - 10^{-2}$ for weak intra-cluster shocks with sonic Mach number $\mathcal{M}_s \sim 2.25-5$.
The authors also reported that this level of efficiency agrees with the current upper limits by \textit{Fermi}-LAT.

However, none of these simulations explicitly followed the CRp distribution functions within the simulation. They instead followed either a total energy budget of CRs without explicit information on the spectral slope or modeled a CR spectrum in post-processing \citep[but see, e.g.,][for spectral approaches in galaxies]{Girichidis2020, Hopkins2022, Werhahn2023}.
It is therefore worth revisiting this problem in current state-of-the-art simulations.

For this work, we employed the first simulation of a CR magnetohydrodynamic (MHD) simulation of a 500 $h^{-1}\:c$Mpc constrained cosmological volume with an on-the-fly Fokker-Planck solver to model CRp and electron acceleration (and re-acceleration) at shocks, adiabatic changes, and energy losses due to synchrotron emission and inverse-Compton (IC) scattering off CMB photons.

We presented the results of this simulation related to synchrotron emission from the Cosmic Web in \citet{Boess2023b}, henceforth \synchweb. 
In this work, we focus on the CRp component of the same simulation to study the expected $\gamma$-ray emission from galaxy clusters while accounting for the interplay of CRp acceleration at accretion shocks, their advection within the ICM, and re-acceleration by merger shocks.
This is a pathfinder simulation in the adiabatic limit of structure formation ($3072^3$ particles, 30 million CPUh).

The paper is structured as follows: in Sect. \ref{sec:gammaweb_methods} we describe the simulation and the code modules we employed and our modeling of $\gamma$-ray emission by pion decay from CRp scattering.
Section \ref{sec:gammaweb_allsky} contains our predictions of $\gamma$-ray emission for the full sky via this process.
In Sect. \ref{sec:gammaweb_fermi} we show the $\gamma$ luminosity in comparison to the \textit{Fermi}-LAT limits and compare the spectra of some prominent clusters to observations in Sect. \ref{sec:gammaweb_clusters}.
Section \ref{sec:gammaweb_discussion} contains a discussion of numerical constraints and the choice of CR injection parameters.
 Finally, Sect. \ref{sec:gammaweb_conclusions} contains our conclusions and a summary of our work.
 
\section{Methods}
\label{sec:gammaweb_methods}

\subsection{Initial conditions}

We used the same simulations as in \synchweb.\ Hence, we only present a brief summary of the initial conditions and refer to \citet{Sorce2018} and \citet{Dolag2023}.
These initial conditions are constrained based on galaxy distance modulus and observational redshift, and thus, to the peculiar velocity from the \textsc{CosmicFlows-2} distance modulus survey \citep[][]{Tully2013}.
By using a Wiener filter algorithm \citep[][]{Zaroubi1995, Zaroubi1998} and applying several reconstruction steps \citep[][]{Doumler2013a, Doumler2013b, Doumler2013, Sorce2014, Sorce2015, Sorce2017, SorceTempel2017, SorceTempel2018}, we constructed a density field of the local Universe at the initial redshift using the constrained realization algorithm by \citet{Hoffman1991}.
Usable initial conditions at the desired resolution were then constructed using the software \textsc{Ginnungagap} \footnote{https://code.google.com/p/ginnungagap/}.
The simulation we used (SLOW-CR3072$^3$) is a cosmological box with a side length of $L = 500 h^{-1}\:c$Mpc, centered on the Milky Way.
Throughout this work, we use the Planck cosmology \citep[][]{Planck2014} with matter densities $\Omega_\mathrm{m} = 0.307$ and $\Omega_\mathrm{baryon} = 0.048$, a cosmological constant of $\Omega_\Lambda = 0.692$, and the Hubble parameter $H_0 = 67.77$ km s$^{-1}$ Mpc$^{-1}$.

\subsection{Simulation code}

The simulation was run using \textsc{OpenGadget3} \citep[][]{Groth2023}, a cosmological \textsc{Tree-SPH} code based on \textsc{Gadget2} \citep{Springel2005}. Gravity in \textsc{OpenGadget3} is solved using a Barnes-Hut tree at short range and a particle-mesh (PM) grid for long-range forces.
We employed an updated smoothed particle hydrodynamics (SPH) implementation \citep[][]{Beck2016} with a spatially and time-dependent high-resolution shock-capturing scheme \citep[][]{Dolag2005b, Cullen2010}, a Wendland $C_4$ kernel with 200 neighbors with bias correction \citep[as introduced in][]{Dehnen2012}, and a high-resolution maximum-entropy scheme in the form of thermal conduction \citep[][]{Jubelgas2004, Arth2014}.

The MHD solver was presented in \citet{Dolag2009}.
We also included nonideal MHD in the form of magnetic diffusion and dissipation as introduced in \citet{Bonafede2011}.
Divergence of the magnetic field arising from numerical truncation errors was cleaned via the hyperbolic cleaning scheme introduced by \citet{Tricco2016} (for the implementation in \textsc{OpenGadget3}, see \citet{Steinwandel2025}).
Shock properties needed for CR acceleration were captured on the fly with the algorithm described in \citet[][]{Beck2016a}.

\subsection{Cosmic-ray model}

As in \synchweb, we used the on-the-fly Fokker-Planck solver \textsc{Crescendo} introduced in \citet{Boess2023} to include CRs in our simulation.
\textsc{Crescendo} models the population of CRps and electrons as piecewise power laws in momentum space and evolves their distribution function in time by solving the diffusion-advection equation in the two-moment approach, following \citet{Miniati2001}.
We accounted for injection at shocks following DSA, adiabatic changes due to expansion or contraction of the surrounding thermal gas, and energy losses of electrons due to synchrotron emission and IC scattering off CMB photons.
This reduces the CR transport equation to
\begin{align}
        \frac{D f(p,\mathbf{x},t)}{Dt} &=  \left( \frac{1}{3} \nabla \cdot \mathbf{u} \right) p \frac{\partial f(p,\mathbf{x},t)}{\partial p} \label{eq:gammaweb_fp-adiabatic} \\
        &+ \frac{1}{p^2} \frac{\partial}{\partial p } \left( p^2 \sum_l b_l f(p,\mathbf{x},t) \right) \label{eq:gammaweb_fp-rad}\\
        &+ j(\mathbf{x}, p, t), \label{eq:gammaweb_fp-sources}
\end{align}
where $\sum_l b_l$ is the sum of all radiative loss mechanisms, and $j(\mathbf{x}, p, t)$ is the source term. 
CR diffusion and streaming are not included, as their computational costs are prohibitive at the scale of this simulation.
Further, recent work by \citet{Reichherzer2025} indicated that CRs can be efficiently confined in the ICM, which reduces their effective diffusion coefficient by four orders of magnitude.
Under these assumptions, advection is the dominant transport process of CRs, which is implicitly included due to the Lagrangian nature of our code.
Protons were considered in the dimensionless momentum range $\hat{p} \equiv \frac{p}{m_p c} \in [0.1, 10^5]$.
We discretized the populations with six bins (1/dex).
CR acceleration was modeled following the DSA model by \citet{Ryu2019} and the shock obliquity-dependent efficiency model by \citet{Pais2018}.
The relevant shock properties for the Mach-number-dependent acceleration model were captured on-the-fly via the shock finding method introduced in \citet{Beck2016a}.
Our shock finder uses an inverted kernel weighting along the local pressure gradient to construct up- and downstream quantities of the shock.
The shock properties were subsequently evaluated following the Rankine-Hugoniot jump conditions.
For the shock obliquity-dependent acceleration efficiency, where the shock obliquity is defined as
\begin{equation}
    \theta_B \equiv \frac{\mathbf{\hat{n}}_s \cdot \mathbf{B}_u}{\vert \mathbf{B}_u \vert} \: ,
\end{equation}
we computed the kernel-weighted magnetic field vector upstream of the shock ($\mathbf{B}_u$) and constructed the angle between this vector and the shock normal ($\mathbf{\hat{n}}_s$) via the cosine similarity \citep[we refer to][Sect. 2.7.3 for detailed descriptions, and Sect. 3.3.2 for tests of this method]{Boess2023}.
We injected the fraction of shock energy that is available for CR acceleration into the full momentum range of the CR population as a single power law following a linear DSA slope at a fixed electron-to-proton-energy ratio of $K_\mathrm{ep} = 0.01$.
The fixed injection momentum of $\hat{p}_\mathrm{inj} = 0.1$ is a compromise between typical injection momenta of $\hat{p}_\mathrm{inj} \sim 0.01$ for accretion shocks and $\hat{p}_\mathrm{inj} \sim 1$ for merger shocks in the ICM \citep[see bottom panel of Fig.~8 in][]{Ha2023}.
The contribution of injected CRs to the total pressure was found by solving the energy integral in each momentum bin and assuming the relativistic equation of state to obtain a closure \citep[see Sect. 2.9 in][for details]{Boess2023}.
For this run, we used closed boundary conditions at the lower end of the distribution function, as it provides more numerical stability and mimics low-momentum cooling on adiabatic compression, which was not explicitly included in this simulation.
However, this choice of boundary condition has only a percent-level impact on the $\gamma$-ray emission of the spectrum when compared to a self-consistent treatment of low-momentum cooling (see Appendix \ref{app:boundary} for a discussion).

\subsection{Gamma-ray emission}\label{sec:gammaweb_gamma_emission}

\subsubsection{Gamma-ray emission from a proton spectrum}

We computed the $\gamma$-ray emission directly from the distribution functions of CRps obtained from the on-the-fly Fokker-Planck solver without any further assumptions for gas particles that do not contain a CR population.
For $\gamma$-ray emission in the range 100 MeV to 100 GeV, the decay of $\pi^0$ mesons is the dominating channel, with Bremsstrahlung and IC up-scattering of background photons being subdominant by roughly an order of magnitude \citep[see Fig. 13 in][]{Yang2018}.
To model the $\gamma$-ray emission by $\pi^0$-decay, we closely followed the approach by \citet{Werhahn2021} and computed the source function for a given $\gamma$-photon energy $E_\gamma$ in units of photons GeV$^{-1}$ s$^{-1}$ cm$^{-3}$ as
\begin{equation}
    q_\gamma(E_\gamma) = 4\pi c \: n_H \int\limits_{\hat{p}_\mathrm{thr}}^{\hat{p}_\mathrm{max}} \mrd \hat{p} \:\: \hat{p}^2 \: f(\hat{p}) \: \frac{\mrd\sigma_\gamma(E_\gamma, \hat{p})}{\mrd E_\gamma}
    \label{eq:gammaweb_gamma_source}
,\end{equation}
where $n_H$ is the number density of hydrogen of the gas particle, $\hat{p}_\mathrm{thr}$ is the threshold momentum for $\pi^0$ production, $\hat{p}_\mathrm{max}$ is the maximum momentum of the distribution function $f(\hat{p})$ in dimensionless momentum space, and
\begin{equation}
    \frac{\mrd\sigma_\gamma(E_\gamma, \hat{p})}{\mrd E} = A_\mathrm{max}(T_p(\hat{p})) \: F(T_p(\hat{p}), E_\gamma)
\end{equation}
is the differential $\gamma$-ray cross section as parameterized in \citet{Kafexhiu2014}.
Here, $E_\gamma$ is the photon energy and $T_p(\hat{p})$ is the kinetic energy of the protons.
$A_\mathrm{max}(T_p(\hat{p}))$ describes the peak of the differential cross section and $F(T_p(\hat{p}), E_\gamma)$ parameterizes the fit to the simulated $\gamma$-ray spectrum.\\
We used their fit to the \textsc{Geant4} data and integrated Eq.~\ref{eq:gammaweb_gamma_source} by solving the contribution of each momentum bin individually, accounting for partially filled bins between $\hat{p}_\mathrm{thr}$ and $\hat{p}_i$.

From Eq. \ref{eq:gammaweb_gamma_source} we obtained an emissivity per photon energy $E_\gamma$ in units [$\gamma$ s$^{-1}$ cm$^{-3}$] from
\begin{equation}
    j_\gamma(E_\gamma) = E_\gamma \: q_\gamma \quad .
\end{equation}
Equation \ref{eq:gammaweb_gamma_source} relates to the total photon flux in units [$\gamma$ s$^{-1}$ cm$^{-2}$] for a given energy band $E \in [E_1, E_2]$ via
\begin{equation}
    F_\gamma = \frac{1}{4\pi d^2} \int_\Omega \mrd V \int\limits_{E_1}^{E_2} \mrd E \: q_\gamma
    \label{eq:gammaweb_gamma_flux}
,\end{equation}
with $d$ being the distance from the emitting region to the observer, and its luminosity in units [erg s$^{-1}$] via
\begin{equation}
    L_\gamma = \int_V \mrd V \int\limits_{E_1}^{E_2} \mrd E \: E \: j_\gamma
    \label{eq:gammaweb_gamma_luminosity}
,\end{equation}
where $V$ is the volume of the emitting region, in our case, the volume of an SPH gas particle with a CRp population.
The emissivity per particle across the whole observational band is found by omitting the volume integral in Eq. \ref{eq:gammaweb_gamma_luminosity}
(for details on the calculation of $q_\gamma$ within our model and tests, see Appendices~\ref{app:gamma_calc} and \ref{app:gamma}, respectively).
We used this method for all calculations of $\gamma$-ray emission unless explicitly stated otherwise.

\subsubsection{Analytic model for gamma-ray emission}

To compare to previous work in the literature, we computed the $\gamma$-ray emission following the analytic model by \citet{Pfrommer2004}.
They derived the source function of $\gamma$-rays from a power-law energy spectrum of protons with slope $\alpha_p$ as 
\begin{align}
    q_\gamma(E_\gamma) \: \mrd E_\gamma \: \mrd V &\approx c \sigma_\mathrm{pp} n_H \Tilde{n}_\mathrm{CR} \frac{2^{4-\alpha_\gamma}}{3\alpha_\gamma} \\
        & \times \left( \frac{m_{\pi^0} c^2}{\mathrm{GeV}} \right)^{-\alpha_\gamma} \left[ \left( \frac{2E_\gamma}{m_{\pi^0} c^2} \right)^{\delta_\gamma} + \left( \frac{2E_\gamma}{m_{\pi^0} c^2} \right)^{-\delta_\gamma} \right]^{\frac{-\alpha_\gamma}{\delta_\gamma}} \\
        &\times \frac{\mrd E_\gamma}{\mathrm{GeV}} \mrd V,
\end{align}
where $\sigma_\mathrm{pp} = 32 \times ( 0.96 + e^{4.4 - 2.4\alpha_\gamma} )$ mbarn is the effective cross section for the pp process, $\alpha_\gamma = 4.3 (\alpha_p - 1/2)$ is the slope of the $\gamma$-ray spectrum, and $\delta_\gamma = 0.14 \alpha_\gamma^{-1.6} + 0.44$ is the shape parameter.
The CRp number density $\Tilde{n}_\mathrm{CR}$ relates to the thermal energy density $\epsilon_\mathrm{th}$ via Eq.~8 in \citet{Pfrommer2004} as 
\begin{equation}
    \Tilde{n}_\mathrm{CR} = X_\mathrm{CR} \epsilon_\mathrm{th} \frac{2(\alpha_p - 1)}{m_p c^2} \left( \frac{m_p c^2}{\mathrm{GeV}} \right)^{\alpha_p - 1} \left[ \mathcal{B}\left( \frac{\alpha_p - 2}{2}, \frac{3 - \alpha_p}{2} \right) \right]^{-1}
,\end{equation}
where $\mathcal{B}(x,y)$ is the beta function, and $X_\mathrm{CR} = \frac{P_\mathrm{CR}}{P_\mathrm{th}}$ is the CRp to thermal pressure ratio within one resolution element.
We used this method as a reference for the diffuse emission in the Coma, Virgo, and Perseus clusters and the integrated $\gamma$-ray spectra in Sect. \ref{sec:gammaweb_clusters}.

% allsky predictions
\section{Full-sky projections\label{sec:gammaweb_allsky}}
\begin{figure*}[h]
\thisfloatsetup{
  floatwidth=\textwidth,
  capbesideposition={right,center},
  capbesidewidth=0.28\textwidth
}
\floatbox[\capbeside]{figure}[\FBwidth]%
{\caption{Full-sky Mollweide projection of the simulation in Galactic coordinates. \textit{Top}: CRp pressure component as the mean value along the line of sight between $r = 10 - 300$ Mpc. This shows the total injected proton component with adiabatic compression as it settles into the higher-density regions of clusters and filaments. The circles indicate the projected $r_\mathrm{vir}$ of each of the labeled cluster match. \textit{Bottom}: Integrated $\gamma$-ray intensity obtained by integrating over the CRp spectra according to Eq. \ref{eq:gammaweb_gamma_luminosity} in the \textit{Fermi}-LAT band $E_\gamma \in [0.5-200]$ GeV.}
 \label{fig:gammaweb_allsky}
}
{%
  \begin{minipage}[t]{0.55\textwidth}
    \includegraphics[width=\linewidth]{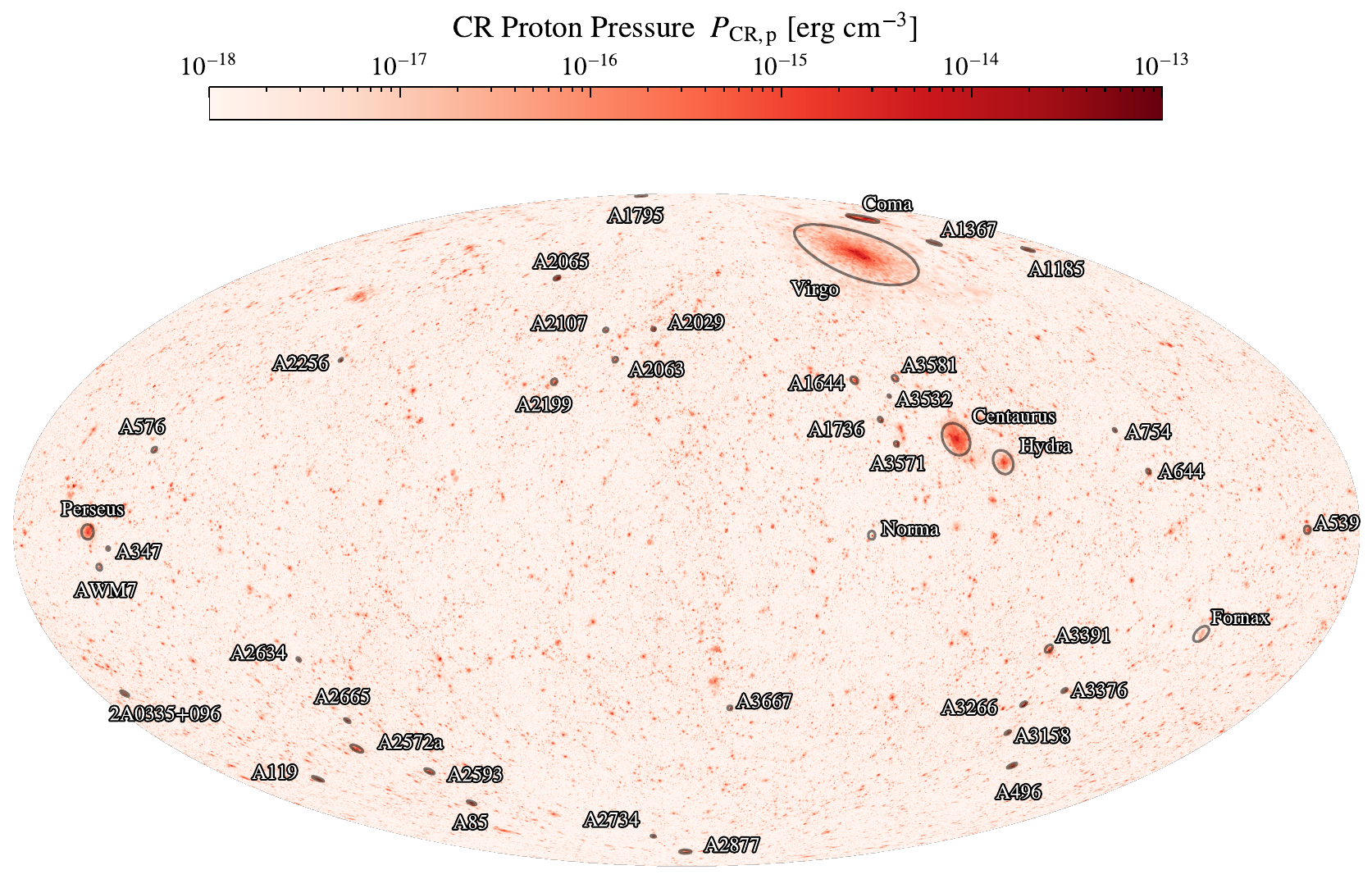}\par

    \vspace{0.6\baselineskip}
    \includegraphics[width=\linewidth]{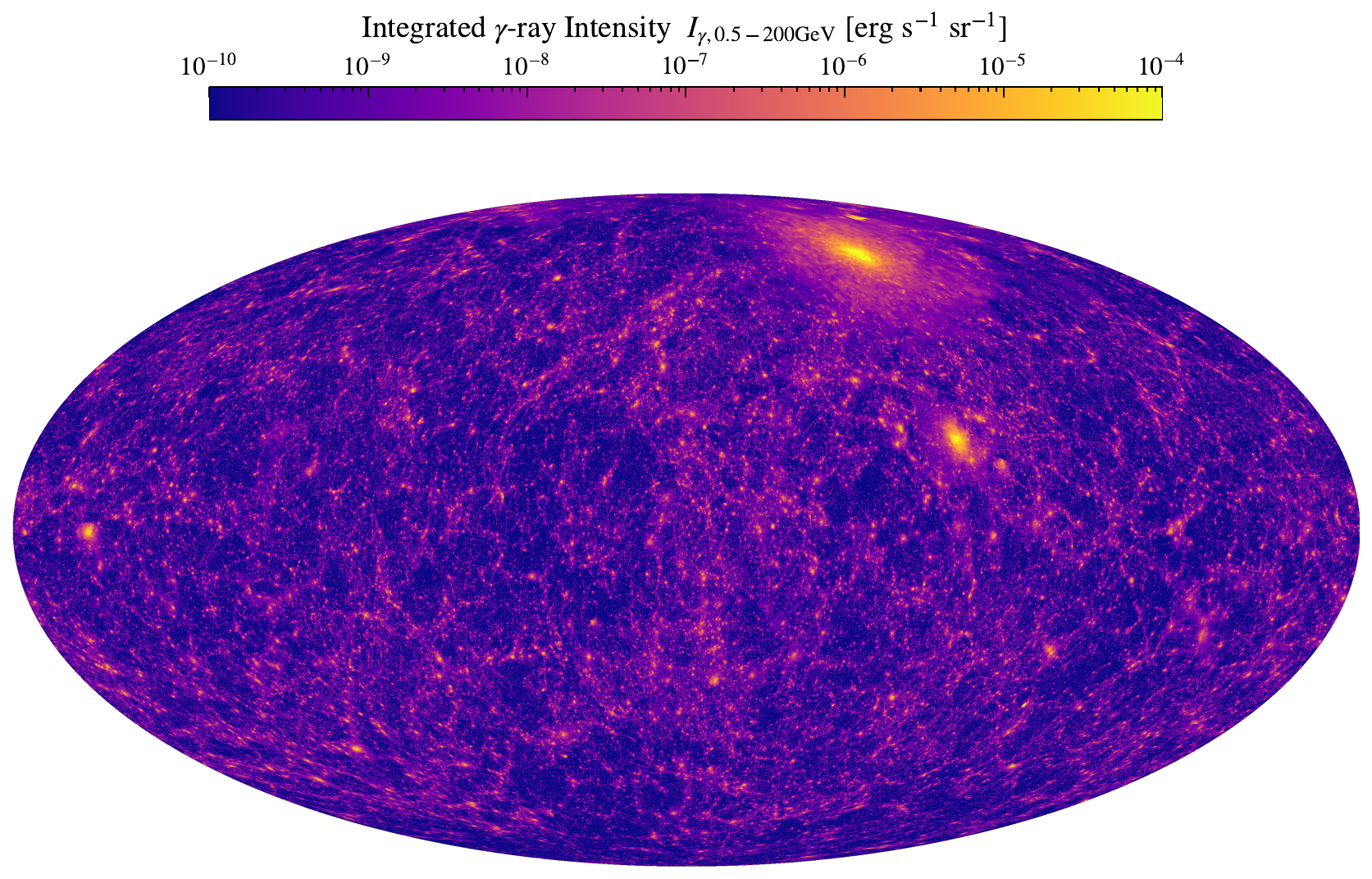}
  \end{minipage}%
}
\end{figure*}
As in \synchweb, we first focused on the morphological analysis of the results.
To do this, we again mapped the simulated SPH quantities on a \textsc{HealPix} sphere \citep[][]{Gorski2005, Dolag2005}.
We chose the same observer view as in \citet{Dolag2023} to match the position of the simulated and observed clusters in our full-sky projection as closely as possible.
Full-sky projections are performed for the output at $z=0$.

\subsection{Cosmic-ray protons}
 Cosmic-ray protons loose hardly any energy in the environments of galaxy clusters and cosmic web filaments.
The main energy-loss mechanisms for the overall energy density of CRps are Coulomb losses due to interaction with the background thermal gas and Alfvén streaming losses \citep[see, e.g.,][]{Pfrommer2004}.
In galaxy clusters, the cooling time of low-momentum protons is on the order of the Hubble time \citep[e.g.,][]{Blasi2007} and therefore of little importance to the overall evolution of the proton spectra.
This holds even more in the extremely low-density environment of cosmic web filaments.

In Fig. \ref{fig:gammaweb_allsky} we show the mean CRp pressure along the line of sight from the center to a radial distance of 300 Mpc within the simulation.
We cut at this radius to avoid projection effects from the box corners.
The labels indicate the cross-identified clusters from \citet{Hernandez2024}, and the circles indicate the projected virial radius of each cluster.

We find a population of CRps injected by merger and accretion shocks in all cosmic structures.
These protons are advected through the cluster and filament volume, setting into large halos and tracing the cosmic web filaments.
In galaxy clusters, we find populations of CRps that extend up to the virial radius.
Beyond this, most if not all clusters show a faint halo of CRps even far beyond the virial radius, filling the volume between the virial radius and the virial shock, which has been pushed further from the virial radius by internal merger shocks \citep[see][]{Zhang2020a, Zhang2020}.
The same holds for cosmic web filaments, where we find CRp populations closely following the cosmological structure.
As in the case of clusters, we find extended regions of CRps around cosmic web filaments.

\subsection{Gamma-ray emission}
As CRps settle into high-density regions in the cluster center and filaments, they can collide inelastically with the thermal background protons and produce $\pi^0$-ons.
We model their surface brightness in the energy band observable by \textit{Fermi}-LAT $E_\gamma \in [0.5-200]$ GeV in the lower panel of Fig. \ref{fig:gammaweb_allsky}.
This is the integrated intensity, not flux, and it is not corrected for cosmological distance.
This allowed us to estimate the relative brightness between the peak emission in clusters and filaments in general and to disregard the actual distance to the individual clusters.

In general, we find a similar central surface brightness for most of the visible galaxy clusters.
They reach central values of $S_\gamma = 10^{-9} - 10^{-8}$ erg s$^{-1}$ cm$^{-2}$ and decrease by 2-3 orders of magnitude toward $r_\mathrm{vir}$.
The diffuse emission in filaments and around galaxy clusters lies four to five orders of magnitude below the central peak, making detection in this regime highly unlikely in the near future.
We note that under the models we adopted for CR acceleration, these values should be considered as a lower limit on the expected $\gamma$-ray emission.

Similar to the synchrotron radio emission of CR electrons, the effects of shock acceleration are visible.
This is especially evident in the shock in the Perseus cluster, to the left of the map.
We will study the $\gamma$-ray emission from merger shocks in upcoming work using a higher-resolution zoom-in on the initial conditions.

% clusters
\section{\textit{Fermi}-LAT limits}
\label{sec:gammaweb_fermi}
\begin{figure}
        \centering
    \includegraphics[width=8cm]{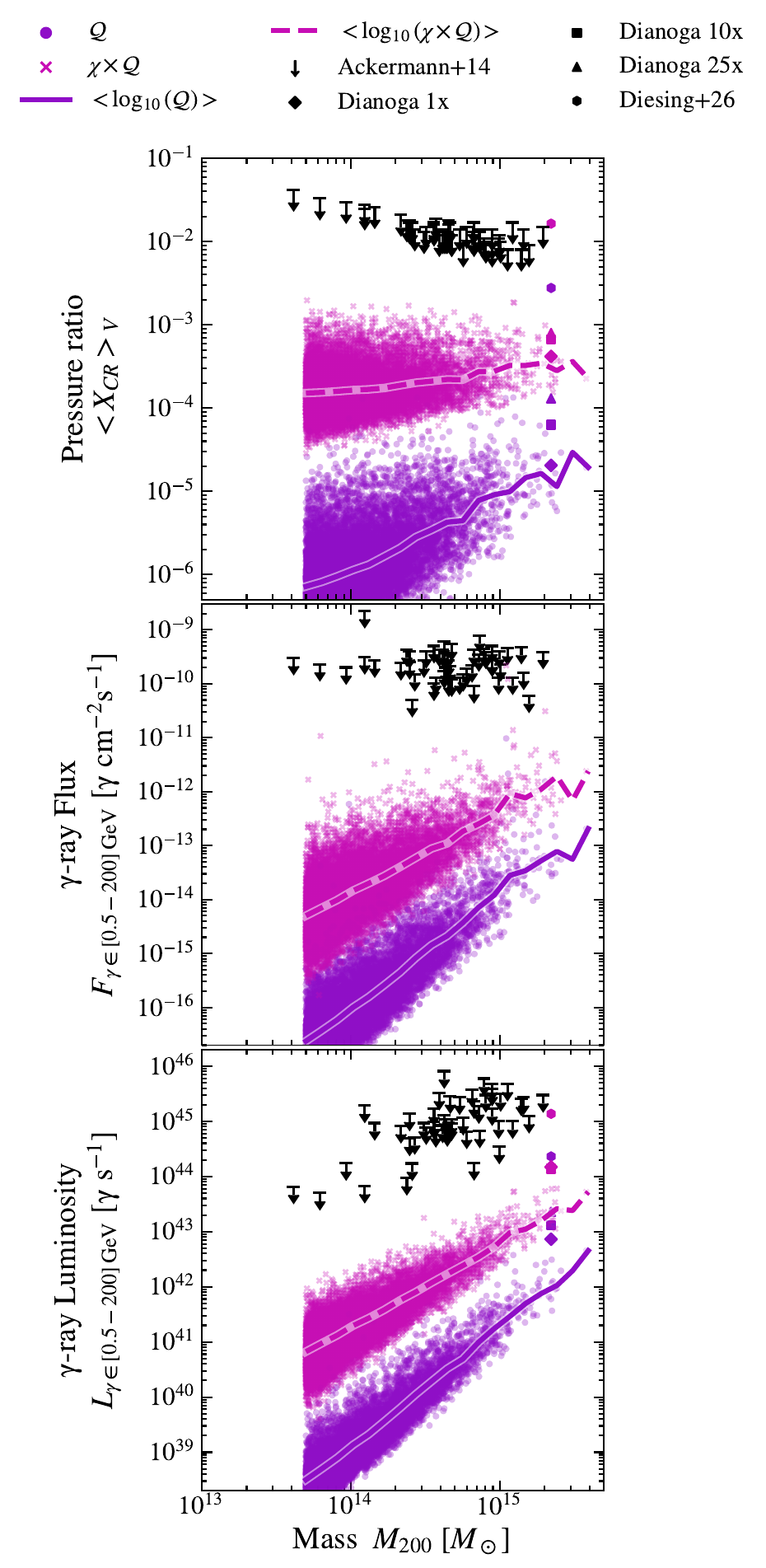}
        \caption{Results for clusters and groups in the simulation box with a virial mass $M_\mathrm{vir} > 5 \times 10^{13} M_\odot$. \textit{Top}: CRp to thermal pressure ratio obtained in the momentum range $\hat{p} \in [0.1, 10^5]$.
 \textit{Middle}: $\gamma$-ray flux obtained at the position of our point of view in Fig.~\ref{fig:gammaweb_allsky} in the energy band $\gamma \in [0.5-200]$ GeV.
 \textit{Bottom}: $\gamma$-ray luminosity in the same energy band.
 In all panels, the dots represent the respective quantity $\mathcal{Q}$ as obtained from the simulation. The crosses indicate the same quantity upscaled by the ratio of $X_\mathrm{CR}$ in the whole cluster volume and $X_\mathrm{CR}$ only in particles that contain a CR population.
 We show the reference values for zoom-in simulations at different resolution levels and varying CR acceleration parameters marked with diamonds, squares, triangles, and hexagons (see Sect.~\ref{sec:gammaweb_shockfinder} for details).
 }
    \label{fig:gammaweb_ackermann}
\end{figure}
\begin{figure}
        \centering
    \resizebox{0.9\hsize}{!}{\includegraphics{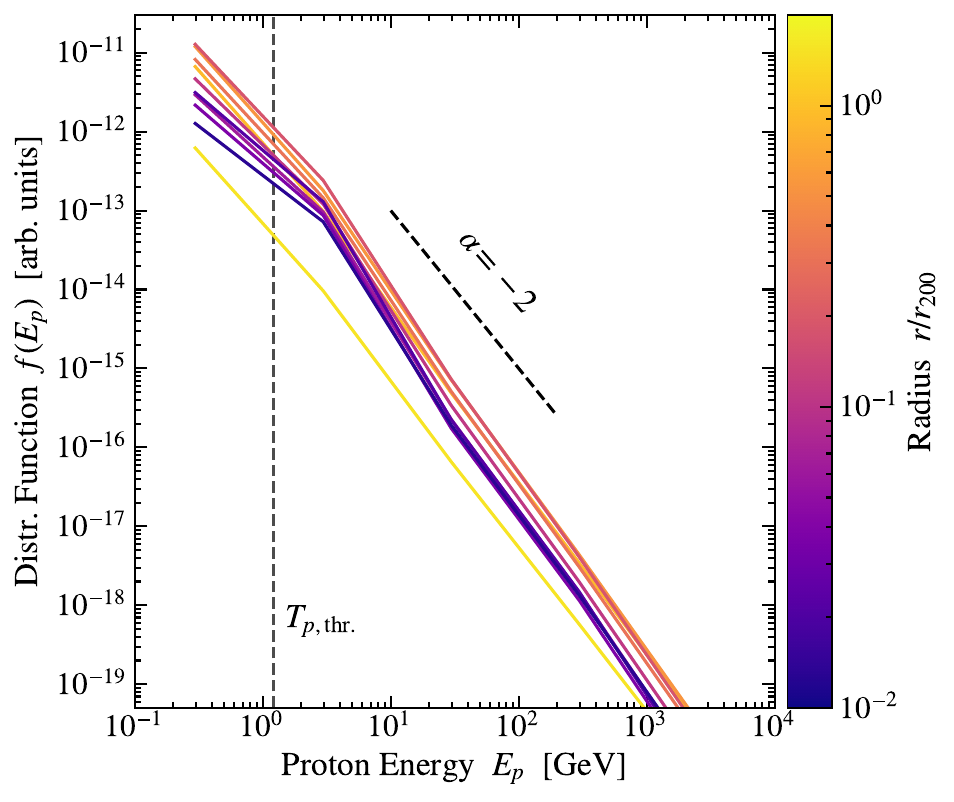}}
        \caption{Mean distribution function in energy space within our Coma cluster replica. We bin the spectra into 20 radial bins indicated by the different colors. The vertical dashed line indicates the threshold energy beyond which protons can be scattered into $\pi^0$-ons.}
    \label{fig:gammaweb_radial_spectra}
\end{figure}
In Fig. \ref{fig:gammaweb_ackermann} we show the comparison of our simulation to the upper limits by \citet{Ackermann2014}.
The top panel shows the ratio of the volume-averaged thermal pressure and CRp pressure $\langle X_\mathrm{CR}\rangle_V = \frac{\langle  P_\mathrm{CR,p}\rangle_V}{\langle P_\mathrm{th}\rangle_V}$.
Within our modeling, gas particles may not contain CR populations for three (numerical) reasons: 
\begin{itemize}
    \item They have passed through an accretion shock before $z=4$ when our CR solver is activated.
    \item They have only experienced shocks with $\mathcal{M}_s < 2.25$ and therefore the acceleration efficiency was zero.
    \item The on-the-fly shock finder did not capture the shock the particle experienced.
\end{itemize}
We therefore defined a rescaling factor $\chi$ as the ratio of these two pressure ratios,
\begin{equation}
    \chi = \frac{\langle X_\mathrm{CR}(P_\mathrm{CR}>0)\rangle_V}{\langle X_\mathrm{CR}\rangle_V}
.\end{equation}
This can be understood as an upper limit within our pure shock-acceleration framework for CRps.
The lines indicate the logarithmic average of the respective values. $P_\mathrm{CR,p}$ can only be considered as a lower limit because of our approach of solving the distribution function in the ultra-relativistic limit.
In the range of $\hat{p} \in [0.1, 1.0)$, this is not an exact approximation and leads to an underestimation of the pressure \citep[see discussion in][]{Girichidis2020}.

For the most massive clusters, the volume-averaged ratio of CRs and thermal pressure lies two to three orders of magnitude below the currently observed limits.
This is driven by only a fraction of the gas in the simulation carrying a CR population, which is a direct result of our model for CR acceleration (we discuss this in more detail in Sect.~\ref{sec:gammaweb_shockfinder}).
When we only consider the average within gas particles that do carry a CR population (crosses in Fig.~\ref{fig:gammaweb_ackermann}), the average pressure ratio is pushed to within one to two orders of magnitude of the current limits.
This agrees with analytic solutions of our employed CR acceleration model \citep[see Fig. 5 in][]{Boess2023} when we take into account that most shocks show a shock obliquity $\theta_B \gg 0^\circ$ (see Fig. 13 in \synchweb).

The middle panel shows the $\gamma$-ray flux within $r_{200}$ as calculated at the same viewpoint used for the full-sky projections in \citet{Dolag2023}.
The flux in the most massive clusters is three to four orders of magnitude below the upper limits found in \citet{Pinzke2011, Ackermann2014}.

To exclude the impact of differences between the simulated and observed distances to clusters, we converted the observed $\gamma$-ray flux into an intrinsic luminosity by using the reported redshifts in \citet{Ackermann2014} \citep[similar to][]{Ha2020}.
Similarly to the pressure ratio, we find a total luminosity that is two to three orders of magnitude below the observational limit.

\
The reason for this discrepancy is twofold: one reason is the numerical behavior of our shock finder.
As discussed in \synchweb, our shock finder is resolution limited at this resolution level.
This means that we do not capture all shocks, especially in the low Mach number regime inside galaxy clusters.
These shocks are expected to contribute significantly to the $\gamma$-ray emission in galaxy clusters, as the protons accelerated at these shocks can be advected through the cluster volume.
Since $\gamma$-ray emission scales linearly with the thermal gas density, observability is biased toward the denser ICM and not the diffuse ICM, where our shock finder performs better.
In addition, due to our shock obliquity-dependent acceleration model and the bias toward high-obliquity shocks due to the alignment of the magnetic field with the shock surface over the shock passage, we artificially suppress proton acceleration (see Fig.~\oblfig~in \synchweb).
We discuss the impact of the shock-finder performance on our findings in more detail in Sect.~\ref{sec:gammaweb_shockfinder}.

The second reason is shown in Fig. \ref{fig:gammaweb_radial_spectra}.
Here, we show the distribution function of the protons in energy space, color-coded by their distance from the center in our simulated Coma cluster replica.
Flat spectra are found in the outskirts of the cluster show with the bright yellow line, approaching a power law with index $\alpha = -2$.
This stems from strong shocks in the cluster periphery, caused by accretion and outward-moving merger shocks.
Farther inside the cluster, shown with increasingly darker lines, two behaviors are visible.
First, the spectra become steeper because these protons have been accelerated by weaker shocks, typically $\mathcal{M}_s \sim 2-4$ in cluster centers \citep[e.g.,][]{Ryu2003, Vazza2011, Kang2013, Skillman2013, Schaal2015, Banfi2020}.
Second, the lowest-energy bin flattens due to adiabatic compression and our closed lower boundary of the spectrum.
This shifts the protons to higher-momentum bins without allowing for an influx of protons from the supra-thermal pool.
With these steeper spectra, a larger fraction of the total energy density resides in the lower-momentum bins.
The lower energy in these bins therefore leads to a significantly lower total pressure.
This is independent of the $\gamma$-ray emission, however, since the threshold energy for $\pi^0$-on production, as indicated by the horizontal dashed line, lies at 1.22 GeV, and with this, beyond the flattened bin.

\section{Coma, Virgo, and Perseus clusters\label{sec:gammaweb_clusters}}
Next, we focused on the $\gamma$-ray emission from the reproductions of Coma, Virgo, and Perseus clusters in our simulation. For a discussion of the match between simulated and observed clusters, see \citet{Hernandez2024}.

\subsection{Surface brightness}
\begin{figure*}[t]
\thisfloatsetup{
  floatwidth=\textwidth,
  capbesideposition={right,center},
  capbesidewidth=0.28\textwidth
}
\floatbox[\capbeside]{figure}[\FBwidth]%
{\caption{Integrated $\gamma$-ray intensity for three prominent clusters. Each image has a width of $3r_\mathrm{vir}$ of the respective cluster. \textit{Top panels}: $\gamma$-ray surface brightness calculated from the proton spectra in the simulation output. \textit{Bottom panels}: $\gamma$-ray surface brightness when applying the modeling by \citet{Pfrommer2004} with a fixed $X_\mathrm{CR} = 0.01$ for every SPH particle.}
    \label{fig:gammaweb_gamma_flux} 
}
{%
  \begin{minipage}[t]{0.6\textwidth}
    \includegraphics[width=\linewidth]{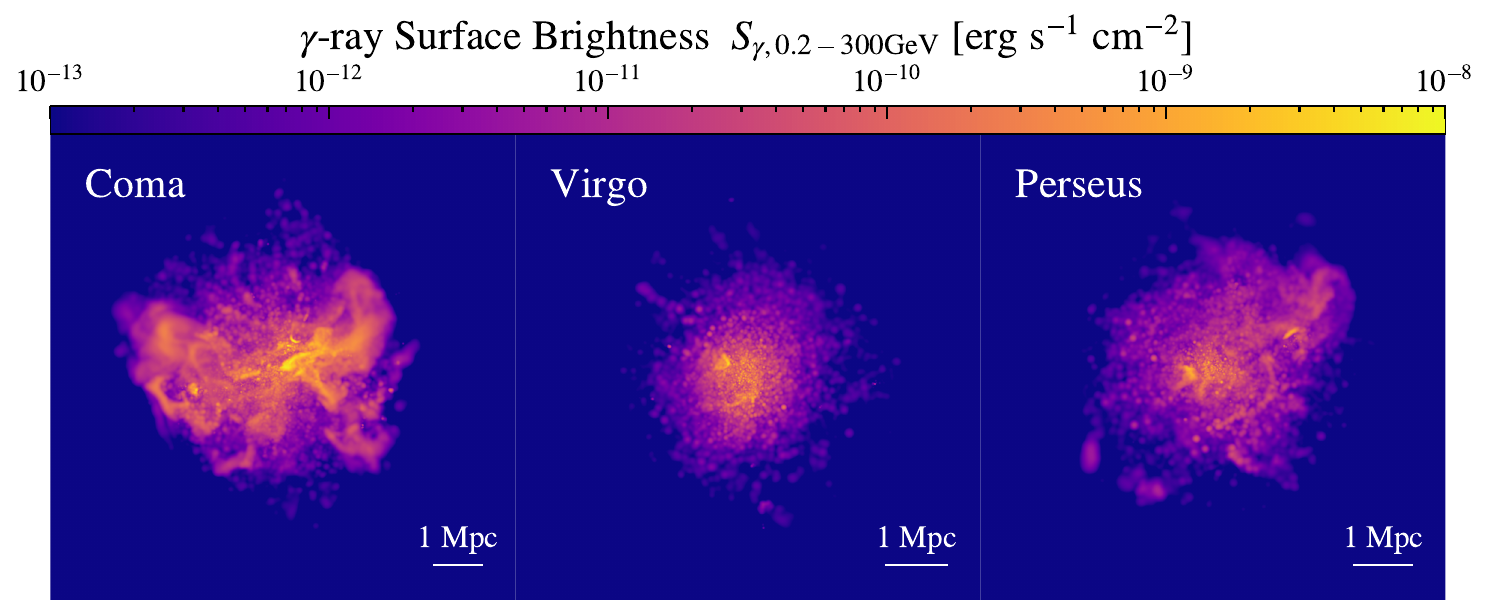}\par

    \vspace{0.2\baselineskip}
    \includegraphics[width=\linewidth]{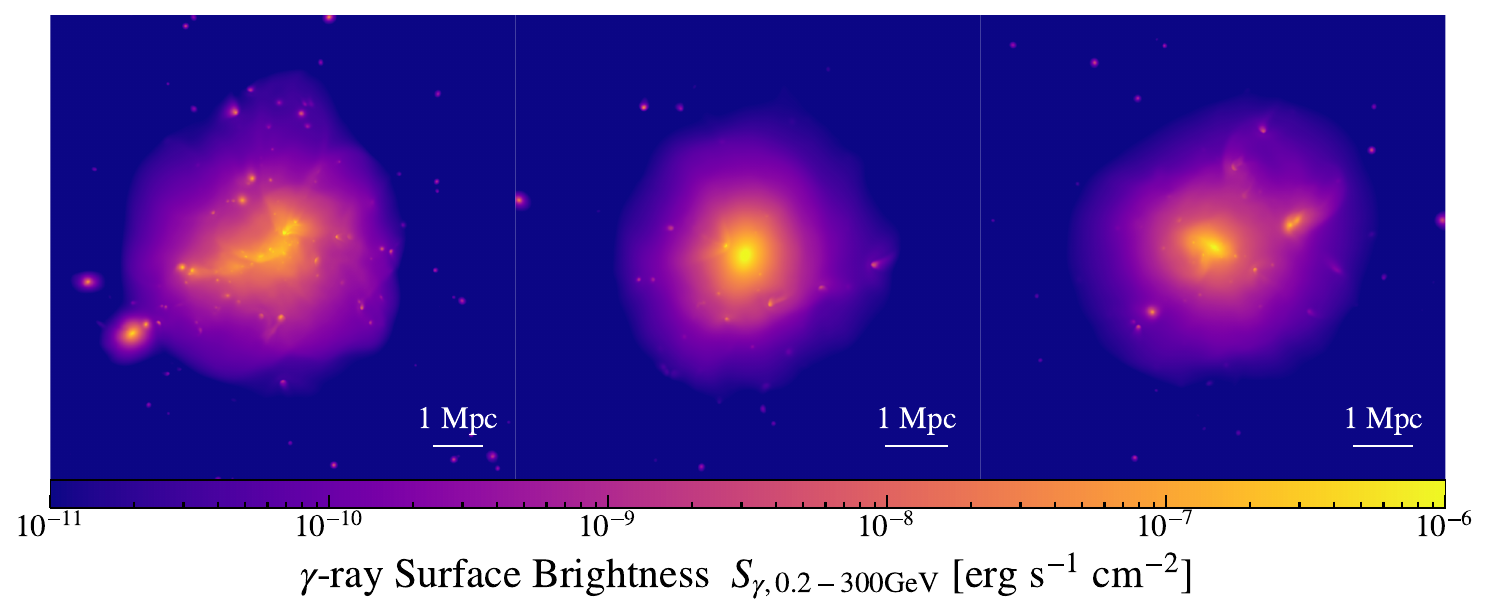}
  \end{minipage}%
}
\end{figure*}
In Fig.~\ref{fig:gammaweb_gamma_flux} we show the $\gamma$-ray surface brightness of our cluster replicas.
We computed the surface brightness by integrating over Eq.~\ref{eq:gammaweb_gamma_source} in the energy range $E_\gamma \in [0.2-300]$ GeV.
Each panel has a width equivalent to $3 r_\mathrm{vir}$ of the respective cluster.
The top panels show the clusters at the resolution of the simulation.
The bottom panels are smoothed with a Gaussian beam corresponding to the single-photon angular resolution at 1 GeV of $\theta = 0.6^\circ$ of the \textit{Fermi}-LAT instrument \citep[][]{Atwood2009}.
We show the size of this angle in the lower left corner of the images for the redshifts of the observed cluster counterparts.

Our Coma cluster is the dynamically most active of the four clusters.
It is in the process of undergoing a merger, driving two shocks in the horizontal direction.
These shocks accelerate CRps and exhibit a $\gamma$-ray surface brightness that is lower by roughly 1.5 orders of magnitude than the central peak.
We also found evidence of CR acceleration at the center, due to small internal shocks, which show the highest surface brightness in Coma.
Due to the angular resolution of \textit{Fermi}-LAT, these structures would not be distinguishable, however, as shown in the bottom panel.

Virgo, on the other hand, is a relatively relaxed cluster in our simulation, with no strong ongoing merger activity \citep[see][for a detailed study of our Virgo replica]{Lebeau2023}. 
This leads to a more strongly peaked central emission, with a steeper decline in surface brightness toward the cluster outskirts.

Our Perseus replica is in the process of a minor merger with a large impact parameter, where a small sub-halo on a highly eccentric orbit drives a shock wave in the Perseus periphery \citep[akin to the main merger geometry in][]{Boess2023a}.
The shock is visible to the top right corner of the image, and the substructure that drives the shock is not visible in the $\gamma$-ray image.
This ongoing merger activity leads to an overall more perturbed cluster and more extended $\gamma$-ray emission than in Virgo.
Nonetheless, we find that all clusters in our set exhibit at least some diffuse $\gamma$-ray emission in their ICM, which is tightly linked with their mass and dynamical state. 

As a reference, we show the surface brightness map obtained using the model by \citet{Pfrommer2004} with $X_\mathrm{CR} = 0.01$ and $\alpha_p = 2.4$ in the bottom panels of Fig.~\ref{fig:gammaweb_gamma_flux}. The color bar is shifted by two orders of magnitude with respect to the upper panels. This figure indicates the two origins of the discrepancy between our results and previous studies of diffuse $\gamma$-ray emission in galaxy clusters: First, the distribution of CRps, and with this, the $\gamma$-ray emission, in our on-the-fly modeling is dominated by shocks and subsequent gas sloshing. Hence, the proton populations scatter off low-density plasma outside of the inner megaparsec of the galaxy cluster. Since the $\gamma$-ray emissivity scales linearly with the background plasma density, this reduces the emissivity.
Second, as shown in Fig.~\ref{fig:gammaweb_ackermann}, the typical values of $X_\mathrm{CR}$ we find in the simulation are considerably below $X_\mathrm{CR} = 0.01$ we assumed for the \citet{Pfrommer2004} model in this figure. We discuss the impact of resolution limitations and our adopted CR acceleration parameters on these results in Sect.~\ref{sec:gammaweb_discussion}.

\subsection{Spectra}
\begin{figure*}[t]
\thisfloatsetup{
  floatwidth=\textwidth,
  capbesideposition={right,center},
  capbesidewidth=0.28\textwidth
}
\floatbox[\capbeside]{figure}[\FBwidth]%
{\caption{Integrated $\gamma$-ray spectra for our three selected clusters: Coma, Virgo, and Perseus. The solid lines show our results, and the arrows indicate the upper limits from observations, where available. The different dashed lines show the results from applying the model by \citet{Pfrommer2004} for a fixed $X_\mathrm{CR} = 0.01$ under varying proton-energy slopes $\alpha_p$.}
    \label{fig:gammaweb_gamma_spectrum}
}
{%
  \begin{minipage}[t]{0.6\textwidth}
    \includegraphics[width=\linewidth]{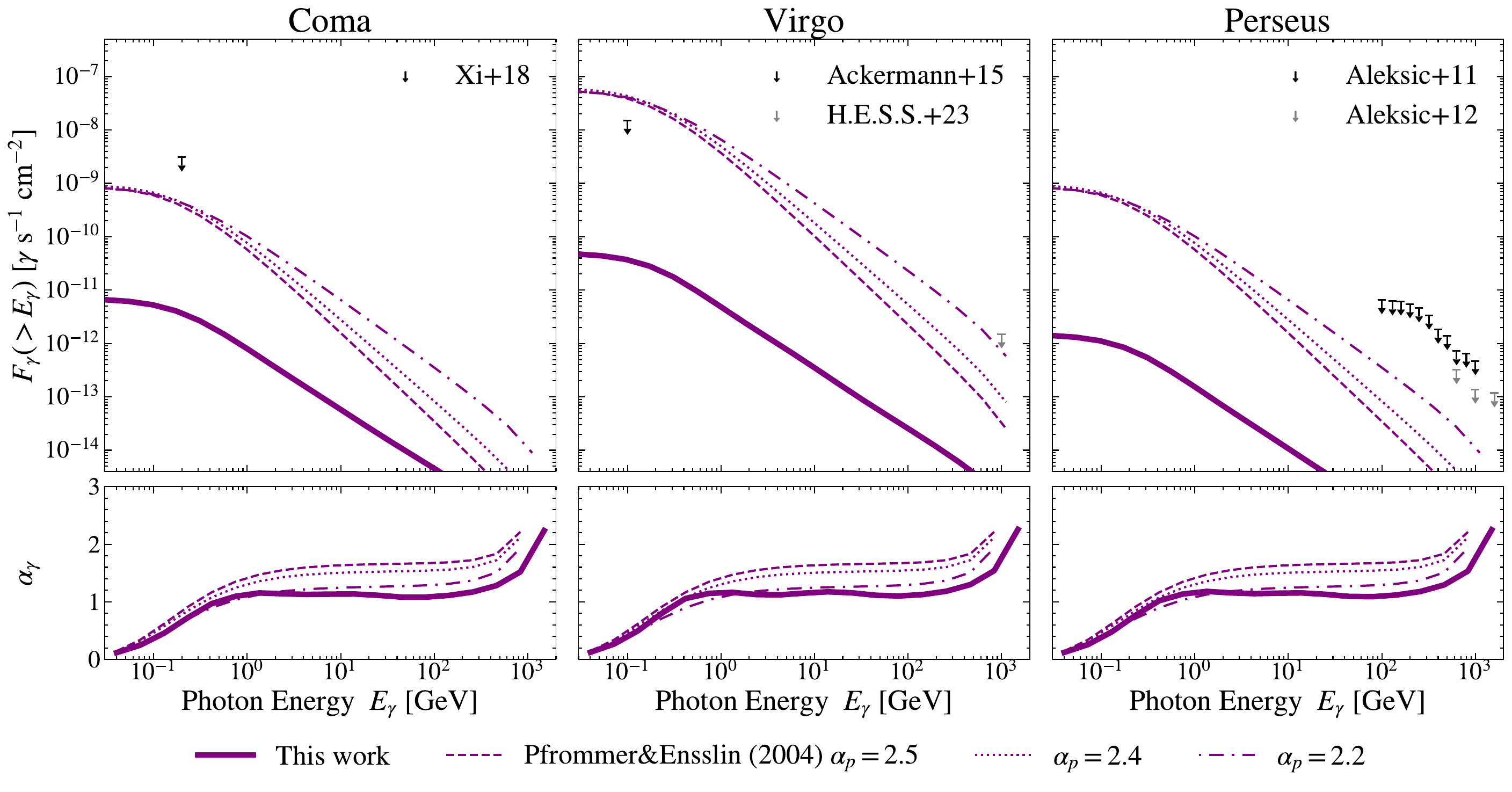}
  \end{minipage}%
}
\end{figure*}
In Fig.~\ref{fig:gammaweb_gamma_spectrum} we show the integrated flux spectra above a threshold energy $F_\gamma(>E_\gamma)$ of our Coma, Virgo, and Perseus cluster replicas.
The top panels show the flux spectrum, obtained by the particles and using the distance between cluster and viewpoint for the full-sky images, and the lower panels show the spectral slope of the $\gamma$-ray spectrum obtained between neighboring photon-energy bins.
Solid lines show our results, and error bars show observational upper limits, where available.
For Coma, we compare to the results by \citet{Xi2018, Adam2021, Baghmanyan2022}.
The upper limits for Virgo were provided by \citet{Ackermann2015, HESS2023} and those for Perseus by \citet{Aleksic2010, Aleksic2012, CTA2023}.

As a reference to the $\gamma$-ray emission calculated from our simulated CRp spectra, the dashed and dotted lines show the model by \citet{Pfrommer2004} with a fixed $X_\mathrm{CR} = 0.01$.
Here, the dashed lines show the result for a proton-energy spectrum with a slope $\alpha_p = 2.5$, the dotted lines show this for a slope with $\alpha_p = 2.4$, and the dash-dotted lines shows this for a slope with $\alpha_p = 2.2$.

In general, the picture is consistent with Fig.~\ref{fig:gammaweb_ackermann}.
Our spectra show fluxes that are lower by two to three orders of magnitude than the observational limits.
Like \citet{Pinzke2011}, we find a universal $\gamma$-ray slope for energies above 1 GeV.
Pinzke et al. reported a spectral slope of the $\gamma$-ray flux that scales as $\alpha_\gamma \approx -1.2$, corresponding to a mean energy slope of $\alpha_p \approx 2.2$.
In our simulation, however, we found a spectral slope of the $\gamma$-ray flux of $\alpha_\gamma \approx -1$, which indicates a mean energy slope for the proton population of $\alpha_p \approx 2.0$.
This is the limit for spectra accelerated by DSA.
We again attribute this to the resolution limitations of our shock finder.
As we showed in \synchweb~and discussed in Sect.~\ref{sec:gammaweb_shockfinder}, the resolution of the cosmological box is not sufficient to capture all shocks with a low Mach number in the ICM.
We are instead biased to stronger shocks, which produce flatter CR spectra.
\section{Discussion\label{sec:gammaweb_discussion}}

\subsection{Impact of the shock-finder accuracy\label{sec:gammaweb_shockfinder}}
\begin{figure*}[t]
    \centering
    \resizebox{\hsize}{!}{\includegraphics{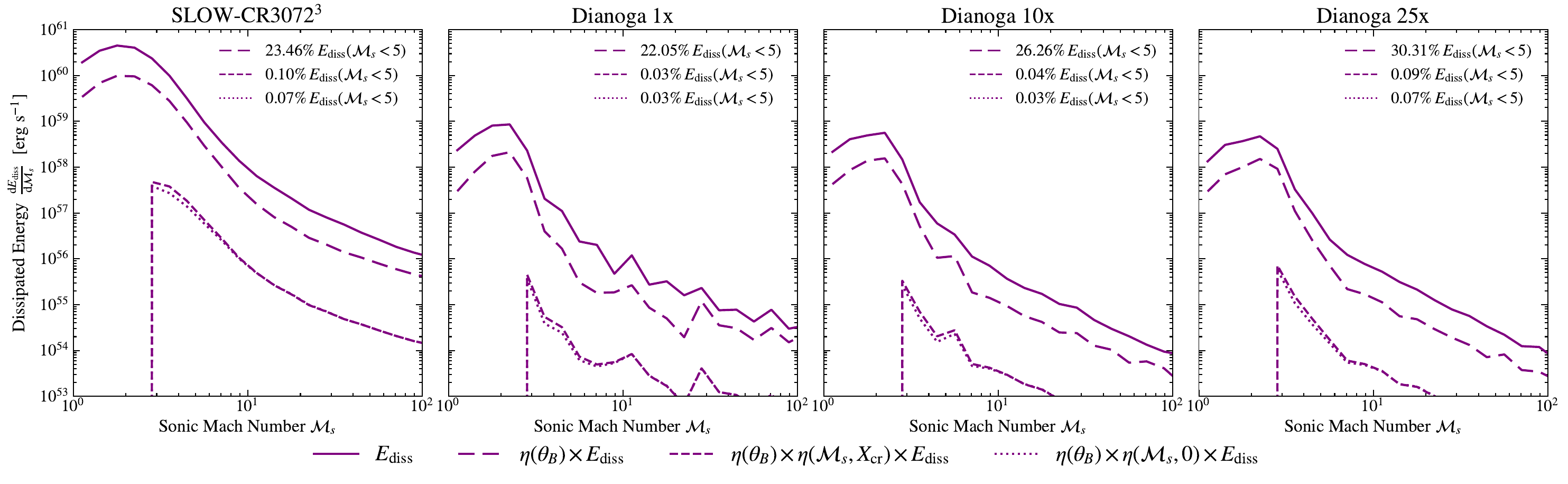}}
    \caption{Histograms of the dissipated energy for the cosmological volume simulation presented in this work (\textit{first panel}) and three reference zoom-in simulations at increasing resolution (\textit{second to fourth panel}). The solid line shows the total kinetic energy flux as identified in the shock finder. The long dashed line shows the energy flux available for CRp acceleration when accounting for the shock obliquity-dependent acceleration model $\eta(\theta_B)$. The short dashed line shows the energy flux available for CR acceleration when accounting for $\eta(\theta_B)$ and re-acceleration of protons as a function of the sonic Mach number. The dotted line shows the same as the dashed line when only accounting for the acceleration of protons from the thermal pool. The legend in each panel indicates the fraction of the total dissipated energy that is available for CRp acceleration under the respective acceleration models.}
    \label{fig:gammaweb_diss_energy}
\end{figure*}

\begin{table}
    \centering
    \caption{Comparison simulations to study the impact of the shock-finder accuracy and acceleration model selection.}
    \resizebox{0.9\columnwidth}{!}{%
    \begin{tabular}{lccc}
    \toprule
        Name & $M_\mathrm{gas}$ [$M_\odot$] & $M_\mathrm{DM}$ [$M_\odot$] & $\epsilon$ [$h^{-1}\,c$kpc] \\
        \midrule
        SLOW-CR3072$^3$     & $8.5 \times 10^7$ & $4.6 \times 10^8$ & 2.7   \\
        Dianoga 1x          & $2.2 \times 10^8$ & $1.2 \times 10^9$ & 11.25 \\
        Dianoga 10x         & $2.2 \times 10^7$ & $1.2 \times 10^8$ & 1.4   \\
        Dianoga 25x         & $8.7 \times 10^6$ & $4.7 \times 10^7$ & 1.0   \\
        \citet{Diesing2026} & $8.7 \times 10^6$ & $4.7 \times 10^7$ & 1.0   \\
    \bottomrule
    \end{tabular}%
    }
    \tablefoot{From left to right, we list the name, mass of the gas particles, mass of the DM particles, and the gravitational softening of the gas particles in the respective simulation.}
    \label{tab:simulations}
\end{table}
Since DSA is the only source of CRs in our simulation, the accuracy of the injected CR population strongly depends on the performance of the shock finder.
To illustrate the impact of the performance of our shock finder, we computed the dissipated energy available for CRp acceleration from the entropy change captured by the hydro solver within the simulation as
\begin{equation}
    E_\mathrm{diss} = \frac{\Delta S}{\gamma - 1} \rho^{\gamma-1} \:,
\end{equation}
where $\Delta S$ is the entropy change over the time step, $\rho$ is the gas density, and $\gamma = 5/3$ is the adiabatic index of the thermal gas.
We used this calculation of the dissipated energy as a basis for the CR acceleration per time step \citep[see][for details]{Boess2023}.
In the left panel of Fig.~\ref{fig:gammaweb_diss_energy}, we show the histogram of dissipated energy as a function of sonic Mach number for the whole simulation domain.
The solid line shows the total dissipated energy.

Similar to \citet{Ryu2003}, \citet{Pfrommer2006}, \citet{Vazza2011}, \citet{Schaal2015}, and \citet{Banfi2020}, we find that the majority of the energy dissipated at the shock lies in the range $\mathcal{M}_s \leq 5$.
The distribution peaks at $\Ms \approx 1.8$, that is, below the critical Mach number required for CR acceleration under the model of \citet{Ryu2019}.

To study the impact of resolution on the shock-finder accuracy, we performed a series of three zoom-in simulations with the same CR injection parameters, but different levels of resolution.
We used initial conditions for the Dianoga simulation set \citep[][]{Bonafede2011}, specifically, the cluster \texttt{g5503149}, which has been used for a number of previous works at various resolution levels \citep[][]{Zhang2020a, Zhang2020, Steinwandel2022_dynamo, Steinwandel2024, Boess2023a, Diesing2026}. We indicate the mass resolution corresponding to 1x, 10x, and 25x in Tab.~\ref{tab:simulations}.\footnote{The resolution of SLOW-CR3072$^3$, the main simulation discussed in this work, corresponds to 3.8x.}
The histograms of dissipated energy in these simulations are shown in panels 2-4 in Fig.~\ref{fig:gammaweb_diss_energy}.
This cluster undergoes a major merger at $z=0$, which drives a shock with Mach number $\mathcal{M}_s \approx 2.5$ and can be seen as dominating the energy dissipation histogram in Fig.~\ref{fig:gammaweb_diss_energy}.

Two effects of an increase in the simulation resolution are observed.
One effect arises because we included a shock-obliquity-dependent efficiency model.
As already discussed in \synchweb, we found shock obliquities that are biased toward high-obliquity shocks (see \synchweb, Fig.~\oblfig) because the shock finder captured a numerically broadened shock.
As the shock propagates through the medium, only the perpendicular component of the magnetic field is compressed, leading to an alignment of the shock surface and magnetic field vector.
In our numerically broadened shock, we also capture the decaying flank of the Mach number distribution, in which the magnetic field is already preferentially aligned with the shock surface.
Together with our description for $\eta(\theta_B)$, this leads to an artificial suppression of CRp accelerations.
The dashed line in the left panel of Fig.~\ref{fig:gammaweb_diss_energy} shows and the legend indicates, that the mean value of $\eta(\theta_B) \sim 0.25$ reduces the available energy for CRp acceleration by 75\%. When we compare this to the lower-resolution Dianoga 1x cluster zoom, the available energy is reduced to 22\%, while at 10x and 25x, the energy is only reduced to 26\% and 30\%, respectively. We therefore conclude that an increase in the resolution of the cosmological box by a factor of ten would increase the energy in CRps by up to 25\%.

The second effect we observe is an increase in $\eta(\mathcal{M}_s)$, as shown by the dashed and dotted lines and indicated in the legend. With an increased numerical resolution, our shock finder is capable of detecting the shocks with more accuracy, and shocks just above the critical Mach number are detected. Between Dianoga 1x and 25x, this increases the energy available for CR acceleration by more than a factor of two. We note, however, that the zoom-in simulations are limited to only one cluster and therefore do not show the same population of shocks as the cosmological volume simulation. Nonetheless, we took the conservative approach to estimate an increase by a factor of two in the total CR energy when the resolution of the cosmological volume simulation increased by a factor of ten.

To summarize, we estimated the combined effect caused by the resolution limitations of our shock finder to be on the order of two to three.
The strong discrepancy between the lower limit we presentend and previous work must therefore stem from the choice of the CR injection parameters.
\subsection{Impact of the CR injection parameters}
As in \synchweb,~we briefly discuss the choice of the CR injection parameters on our results.
For protons, the relevant parameters are the injection momentum $p_\mathrm{inj}$, the injection slope $q_\mathrm{inj}$, and the overall acceleration efficiency $\eta(\mathcal{M}_s, X_\mathrm{CR},\theta_B)$.

\subsubsection{Injection momentum $\hat{p}_\mathrm{inj}$}
We used a fixed injection momentum of $\hat{p}_\mathrm{inj} = 0.1$.
In other recent work \citep[e.g.,][]{Ha2020}, the injection momentum was computed on-the-fly as
\begin{equation}
    \hat{p}_\mathrm{inj,p} = \xi \frac{p_\mathrm{th,p}}{m_p c} = 3.0-3.5 \: \frac{\sqrt{2 k_B m_p T_2}}{m_p c}
    \label{eq:gammaweb_gamma_p_inj}
,\end{equation}
where $T_2$ is the temperature downstream of the shock, and we used $\xi \approx 3.0 - 3.5$ from \citet{Ryu2019}.
Compared to the impact of $p_\mathrm{inj}$ on synchrotron emission, the impact on $\gamma$-ray production is quite significant, as shown in Fig.~\ref{fig:gammaweb_gamma_emissivity}.
We show the emissivity of a CRp population with a fixed total energy in the parameter space of the injection momentum and spectral slope.
The color code indicates the emissivity per parameter space element, and the contours indicate changes by half an order of magnitude.

For strong shocks where $q \sim 4$, the difference between $p_\mathrm{inj} = 0.01$ and $p_\mathrm{inj} = 1$ only leads to an insignificant change in emissivity.
This changes drastically, however, for shocks of $\mathcal{M}_s \approx 3$, and with this, $q = 4.5$.
These weak shocks are expected in the cluster periphery, where they are visible as radio relics \citep[see][for a recent review]{Weeren2019}.
At a slope of $q=4.5$, the emissivity is reduced by two orders of magnitude for an equal reduction in $p_\mathrm{inj}$.
For $\mathcal{M}_s \approx 2$ shocks with $q = 5$, this discrepancy even increases to three orders of magnitude.
As these CRps are accelerated even further in the cluster center and can therefore advect in the cluster volume, this has strong implications for the total $\gamma$-ray emission.
We note, however, that our conservative approach to use a fixed injection momentum of $\hat{p}_\mathrm{inj} = 0.1$ leads to an increased emission compared to typical values of $\hat{p}_\mathrm{inj}$ when computed directly from Eq.~\ref{eq:gammaweb_gamma_p_inj}.

We will study the impact of varying $p_\mathrm{inj}$ in future zoom-in simulations at 8x the current resolution, where we expect our shock finder to be fully converged.

\subsubsection{Injection slope $q_\mathrm{inj}$}

The strong dependence of the $\gamma$-ray emissivity on $q$ also has implications for the parameterization of $q_\mathrm{inj}$.
As discussed in \synchweb, \citet{Caprioli2020} recently reported a magnetosonic post-cursor wave that can efficiently scatter particles from the shock-acceleration region.
The development of this wave depends on the ratio of the downstream Alfvén velocity $v_{A,2}$ and the downstream velocity $u_2$. This modifies $q_\mathrm{inj}$ to
\begin{align}
    q_\mathrm{inj} = \frac{3r}{r - 1 - \alpha}; \:\: \alpha \equiv \frac{v_{A,2}}{u_2}
    \label{eq:gammaweb_gamma_new_dsa_slope},
\end{align}
where $r$ is the shock-compression ratio.

\begin{figure}
    \centering
    \resizebox{0.9\hsize}{!}{\includegraphics{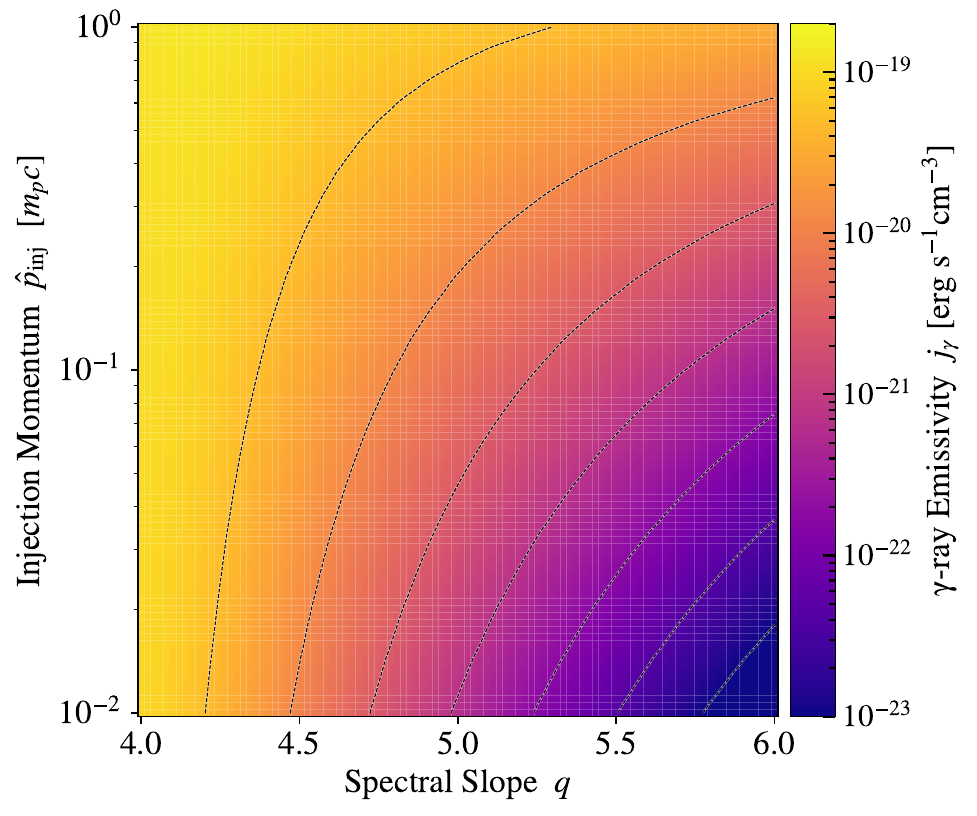}}
    \caption{$\gamma$-ray emissivity as a function of spectral slope $q$ and injection momentum $\hat{p}_\mathrm{inj}$ for a fixed total energy density in CRps. The dashed lines show contours spaced by half an order of magnitude.}
    \label{fig:gammaweb_gamma_emissivity}
\end{figure}

For the low Mach number shocks in the ICM, even low values for $\alpha$ with $\alpha \leq 0.2$ can have a significant impact on the injection slope.
As shown in Fig.~\ref{fig:gammaweb_gamma_emissivity} for our fixed value of $\hat{p}_\mathrm{inj} = 0.1$, a steepening of the spectrum by 0.5 or 1 leads to a decrease of half an order or one order of magnitude in emissivity, respectively.
However, this parameterization was found for low-$\beta$ supernova shocks, which expand into a medium that is vastly different than the high-$\beta$ ICM considered here.
There is currently no work in the literature that explores the dependence $q_\mathrm{inj}(\alpha)$.

\subsubsection{Acceleration efficiency $\eta(\mathcal{M}_s, X_\mathrm{CR}, \theta_B)$}

We employed the DSA parameterization of \citet{Ryu2019}, which shows a weak dependence on the sonic Mach number $\mathcal{M}_s$ for CRp acceleration from the thermal pool.
The re-acceleration efficiency is slightly higher, as shown in Fig.~\ref{fig:gammaweb_diss_energy}, for instance. Our injection model interpolates between acceleration and re-acceleration based on the value of $X_\mathrm{CR}$ currently in the SPH particle and the reference value for $X_\mathrm{CR}$ in the model, however, which in the case of \citet{Ryu2019} is $X_\mathrm{CR} = 0.05$.
Since the initial acceleration from the thermal pool is so inefficient for low Mach number shocks, we rarely reach these values, as shown in the top panel of Fig.~\ref{fig:gammaweb_ackermann}.
Our model is therefore driven by the initial acceleration and not by re-acceleration.
However, the \citet{Ryu2019} model only allows for acceleration of particles beyond $\mathcal{M}_s = 2.25$, which narrows the range of available shocks to accelerate CRs significantly.

 \citet{Aleksic2010, Pinzke2010} used efficiency models that reached 50\% at strong shocks and decreased to 4\% at $\mathcal{M}_s \approx 1.7$ \citep[][]{Ensslin2007}.
Compared to our parameterization of the \citet{Ryu2019} model, this is higher by a factor of 20 for strong shocks and higher by almost one order of magnitude for weak shocks.
Together with the shock-obliquity-dependent acceleration model, which is absent in these works, this can account for a discrepancy of two orders of magnitude in $\gamma$-ray emission from CRps accelerated at weak shocks in galaxy clusters.

This is illustrated in the captions of Fig.~\ref{fig:gammaweb_diss_energy}. Within our modeling, only $0.07\%$ of the dissipated energy of shocks with a Mach number $\mathcal{M}_s < 5$ typical for merger shocks is available for proton acceleration from the thermal pool. This difference between the effective CR acceleration efficiency in our work and the previous work of $\gamma$-ray modeling in galaxy clusters can account for the almost two orders of magnitude lower prediction for diffuse emission presented here.

To emphasize this, we also show results taken from the simulation used in \citet{Diesing2026} in Fig.~\ref{fig:gammaweb_ackermann}, marked as hexagons. In this simulation, we allowed for CR acceleration at shocks with $\mathcal{M}_s > 2$ and always used 10\% of the dissipated energy for CRp acceleration.
This choice of parameters pushes the predicted $\gamma$-ray emission within a factor of one to five of the observed limits by \citet{Ackermann2014}.

Recent simulations by \citet{Orusa2025} indicated that in 3D, quasi-perpendicular shocks can also accelerate CRps, unlike in our modeling of shock-obliquity-dependent acceleration based on \citet{Caprioli2014, Pais2018}.
This might increase the $\gamma$-ray emission in our simulation by another factor of four, as indicated in the legend of Fig.~\ref{fig:gammaweb_diss_energy}.

\subsection{Impact of cosmic-ray transport}
The nature and propagation velocity of CR transport in the ICM are still strongly under debate.
The two most common flavors of CR transport are CR diffusion, where CRs propagate along magnetic field lines with a constant or energy-dependent diffusion coefficient, or CR streaming, where CRs stream down a CR pressure gradient with a streaming velocity $v_\mathrm{st}$ that generally depends on the magnetic field strength and density in the medium and is on the order of the Alfvén or sound speed \citep[see, e.g.,][for a detailed discussion]{Ensslin2011}.
With the long cooling times of CRps, it is expected that regardless of the exact nature of these transport processes, CRps will fill the entire cluster volume over the Hubble time.

The inclusion of CR transport in simulations adds a significant numerical challenge and cost \citep[see][for a recent review]{Hanasz2021}, which made it unfeasible for this large cosmological simulation.
Due to the Lagrangian nature of our simulation code, however, we automatically accounted for the advection of CRs, which is expected to be the dominant transport process for low-energy CRps \citep[][]{Ensslin2011, Wiener2013, Wiener2018, Reichherzer2025}.

We expect that an inclusion of streaming might reduce the $\gamma$-ray flux from the center of the cluster, as CRps would stream down the CR pressure gradient toward the cluster periphery.

The inclusion of diffusion might affect the expected $\gamma$-ray emission because the diffusion coefficient depends on the energy.
The assumption that higher-energy protons diffuse faster/easier would lead to a flattening of the CRp spectrum away from acceleration zones \citep[see, e.g.,][for a study of the evolution of CRp spectra under anisotropic, energy-dependent diffusion]{Girichidis2020}.
This would naturally flatten the $\gamma$-ray spectrum as well.
However, full cosmological simulations of galaxy clusters, including anisotropic and energy-dependent diffusion, are not computationally feasible so far.
This might nonetheless hold interesting prospects for future work.
% conclusions
\section{Conclusions\label{sec:gammaweb_conclusions}}

We showed results from the first constrained cosmological simulation with an on-the-fly spectral CR model to study the time evolution of CRps in the local cosmic web.
Our results are summarized below.
\begin{itemize}
    \item We found CRp acceleration at structure formation and accretion shocks in and around galaxy clusters and cosmic web filaments.
    \item The CRps advect through the cluster and filament volume and settle into halos extending to the virial radius of clusters.
    \item These protons produce volume-filling diffuse $\gamma$-ray emission on cluster scales through hadronic interactions.
    \item Strong shocks produce $\gamma$-ray emission with a surface brightnesses lower by roughly one order of magnitude than the central surface brightness.
    \item In general, we found values for the CR to thermal pressure ratio ($X_\mathrm{CR}$) that are lower by three orders of magnitude than current estimates when averaging over the full cluster volume. This is mainly driven by our choice of the CR acceleration model. The combination of shock-obliquity-dependent acceleration with the latest models based on PIC simulations only allowed a significantly smaller fraction of dissipated shock energy for CR acceleration than was assumed in previous work.
    \item This discrepancy in $X_\mathrm{CR}$ means that our predictions for diffuse $\gamma$-ray flux and luminosity lie three to five orders of magnitude below the current \textit{Fermi}-LAT limits.
\end{itemize}
Future work will have to vary the CR acceleration parameters to provide a more conclusive prediction for diffuse $\gamma$-ray emission from galaxy clusters.

\begin{acknowledgements}
We thank the anonymous referee for their detailed comments, which improved the quality of this manuscript.
LMB would like to thank Maria Werhahn, Damiano Caprioli, Sruthiranjani Ravikularaman, and Jonathan Biteau for helpful discussions.
LMB, IK, DK, KD, and EH acknowledge support by the Excellence Cluster ORIGINS, which is funded by the Deutsche Forschungsgemeinschaft (DFG, German Research Foundation) under Germany's Excellence Strategy - EXC-2094-390783311 and the COMPLEX project from the European Research Council (ERC) under the European Union’s Horizon 2020 research and innovation program grant agreement ERC-2019-AdG 882679.
LMB is supported by NASA through grant 80NSSC24K0173. 
The simulations used in this work were carried out at the Leibniz Supercomputer Center (LRZ) under the project \textit{pn68na}.
The analysis has been carried out on the computing facilities of the Computational Center for Particle and Astrophysics (C2PAP) and of the Leibniz Supercomputer Center (LRZ).

\end{acknowledgements}



\bibliographystyle{aa}
\bibliography{example} % if your bibtex file is called example.bib

%%%%%%%%%%%%%%%%%%%%%%%%%%%%%%%%%%%%%%%%%%%%%%%%%%

%%%%%%%%%%%%%%%%% APPENDICES %%%%%%%%%%%%%%%%%%%%%

\appendix

\section{Gamma-ray emission from CR momentum spectra\label{app:gamma_calc}}

As mentioned in Sect.~\ref{sec:gammaweb_gamma_emission}, our computation of the $\gamma$-ray emissivity closely follows the method described in \cite{Werhahn2021}; see their Appendix 1.
With the difference that we only use the parameterization by \cite{Kafexhiu2014}, and omit the inclusion of \citet{Yang2018} for low CR energies.
We will summarize the core concepts here, for completeness.

The emissivity of a momentum distribution of a CRp spectrum $f(\hat{p})$  is given in Eq.~\ref{eq:gammaweb_gamma_source} and repeated here:
\begin{equation}
    q_\gamma(E_\gamma) = 4\pi c \: n_H \int\limits_{\hat{p}_\mathrm{thr}}^{\hat{p}_\mathrm{max}} \mrd \hat{p} \:\: \hat{p}^2 \: f(\hat{p}) \: \frac{\mrd\sigma_\gamma(E_\gamma, \hat{p})}{\mrd E_\gamma}
    \label{eq:gammaweb_gamma_source_app}
.\end{equation}
The central task of this computation is the differential cross section of pion production, which takes the general form
\begin{equation}
    \frac{\mathrm{d}\sigma_\gamma(E_\gamma, E_p)}{\mathrm{d}E_\gamma} = 2 \int_{E_{\pi,\mathrm{min}}}^{E_{\pi,\mathrm{max}}} \mathrm{d}E_\pi \frac{\mathrm{d}\sigma_\pi(E_p, E_\pi)}{\mathrm{d}E_\pi} f_{\gamma,\pi}(E_\gamma,E_\pi)
    \label{eq:diff_cross_section}
,\end{equation}
where the normalized energy distribution $f_{\gamma,\pi}(E_\gamma,E_\pi)$ is a Green's function which describes the probability for a pion at energy $E_\pi$ to produce a $\gamma$ photon at energy $E_\gamma$, where
\begin{equation}
    f_{\gamma,\pi}(E_\gamma,E_\pi) = \frac{1}{\sqrt{E_\pi^2 - E_{\pi,0}^2}}
    \label{eq:f_gamma_pi}
\end{equation}
following \citet{Stecker1971}.
In the absence of an analytic solution for Eq.~\ref{eq:f_gamma_pi}, \citet{Kafexhiu2014} introduce the parameterization
\begin{equation}
    \frac{\mrd\sigma_\gamma(E_\gamma, \hat{p})}{\mrd E} = A_\mathrm{max}(T_p(\hat{p})) \: F(T_p(\hat{p}), E_\gamma)
    \label{eq:K14_fit}
.\end{equation}
The peak of the differential cross section $A_\mathrm{max}(E_p)$ can be expressed as
\begin{align}
    A_\mathrm{max}(E_p) = \begin{cases}
        b_0 \times \frac{\sigma_\mathrm{pp,inel}(E_p)}{E_\pi^\mathrm{max}},
            &\text{for } E_p^\mathrm{th} \leq E_p \leq 1 \text{ GeV}\\\\
        b_1 \: \theta_p^{-b_2} \: e^{(b_3 \log^2(\theta_p))} \\\times \frac{\sigma_\mathrm{pp,inel}(E_p)}{E_{p,0}},
            & \text{for } E_p \geq 1 \text{ GeV,}
        \end{cases}
\end{align}
where $b_i$ are fit parameters to \textsc{Geant4} data, $E_\pi^\mathrm{max}$ is the maximum pion energy in the lab frame allowed by the participating particle's kinematics, and $E_{p,0}$ is the proton rest mass.
$\sigma_\mathrm{pp,inel}(E_p)$ is the inclusive $\pi^0$ production cross section given as
\begin{align}
    \sigma_\mathrm{pp,inel}(T_p) = &\left[ 30.7 - 0.96\log \left( \frac{T_p}{T_{p,\mathrm{th}}} \right) + 0.18\log^2\left(\frac{T_p}{T_{p,\mathrm{th}}}\right) \right] \\
        \times &\left[ 1 - \left(\frac{T_p}{T_{p,\mathrm{th}}}\right)^{1.9} \right]^3 \mathrm{ mb,}
        \label{eq:sigma_pp_inel}
\end{align}
where $T_p$ is the kinetic energy of the proton in the lab frame and $T_{p,\mathrm{th}} = 2m_\pi c^2 + m_\pi^2 c^4 / (2m_p c^2) \approx 0.2797$ GeV is the threshold proton kinetic energy in the particle lab frame.

The fit formula $ F(E_p, E)$ in Eq.~\ref{eq:K14_fit} can be written as
\begin{equation}
     F(E_p, E_\gamma) = \frac{\left(1 - X_\gamma^{\alpha(E_p)}\right)^{\beta(E_p)}}{\left( 1 + \frac{X_\gamma}{C} \right)^{\gamma(E_p)}}
,\end{equation}
where $\alpha(E_p)$, $\beta(E_p)$, and $\gamma(E_p)$ are fit functions, which we will omit in the interest of brevity, that depend on the proton energy in the laboratory frame $E_p$.
$X_\gamma$ is a ratio between the energy of the emitted $\gamma$-photon, the maximum allowed $\gamma$-photon energy and the rest-mass of a pion.

\section{Tests for $\gamma$-emission\label{app:gamma}}
\begin{figure}
    \centering
    \includegraphics[width=\textwidth]{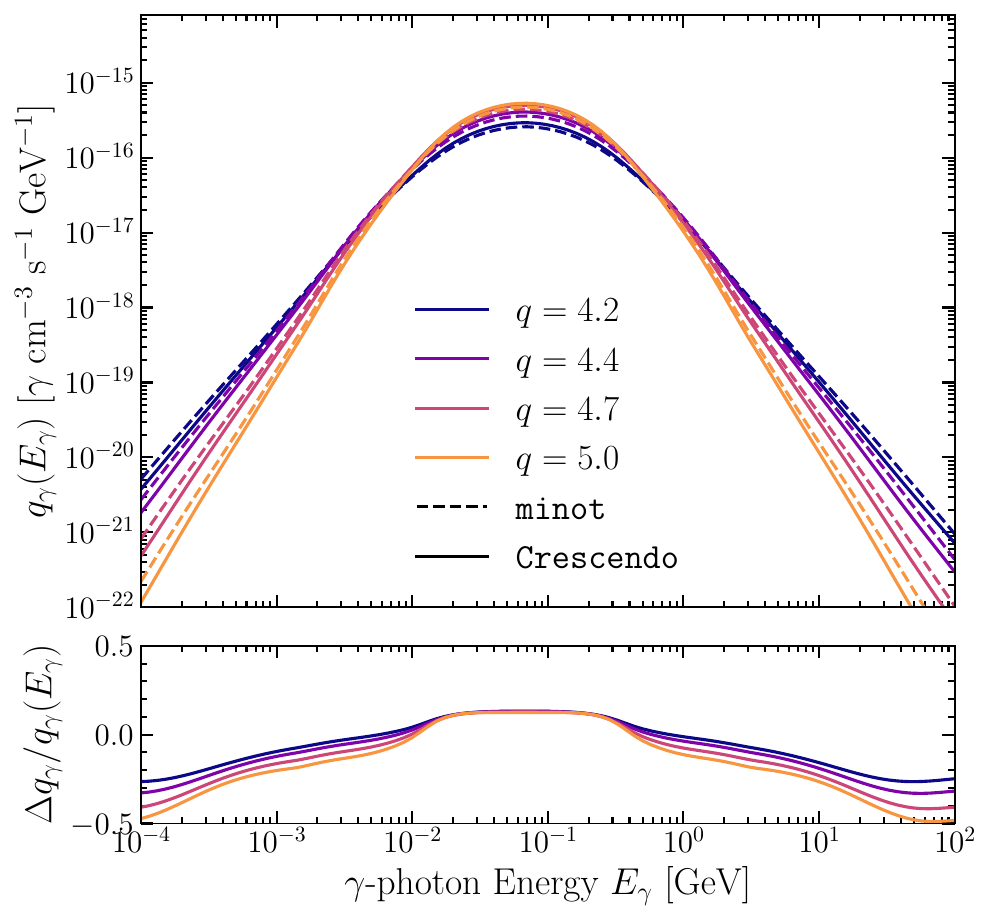}
    
    \includegraphics[width=\textwidth]{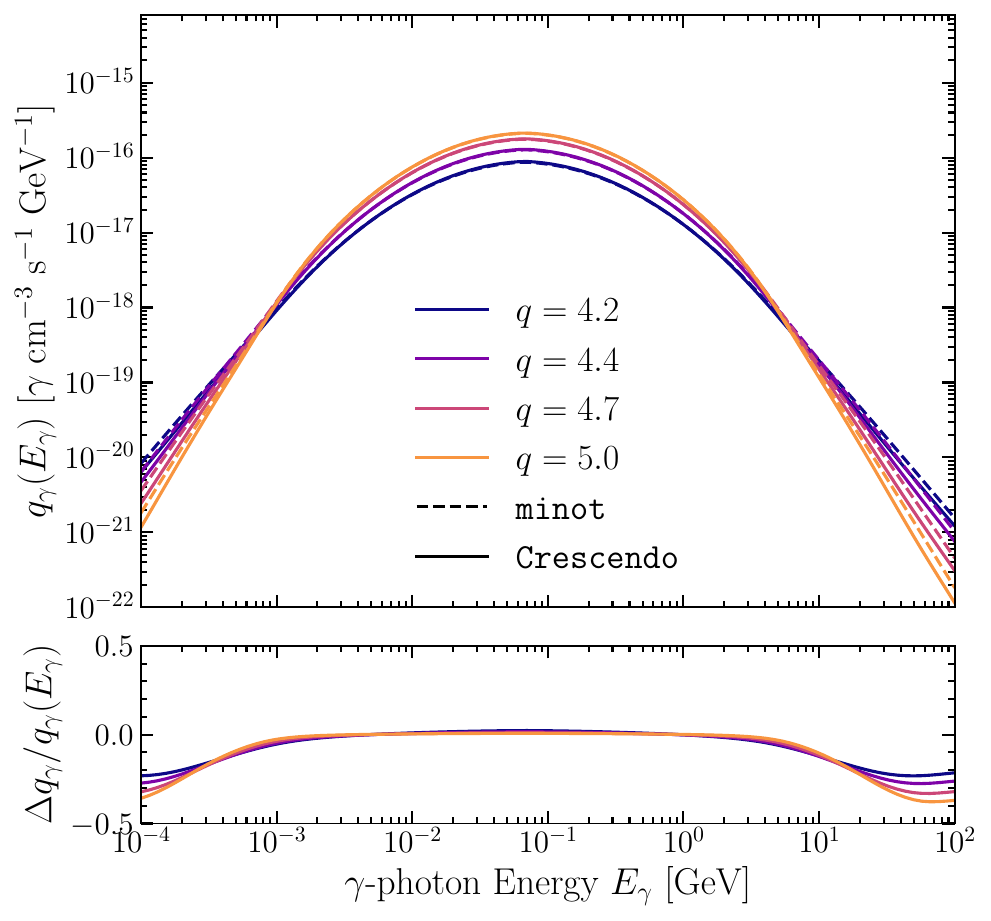}
    \caption{Our calculations of the $\gamma$-ray source function compared to the one by the \textsc{minot} package. For both plots we set up a power law in momentum space with an energy density of $\epsilon_{\mathrm{CR},p} = 1$ erg cm$^{-3}$ and vary the minimum momentum of the power law, $p_\mathrm{min}$.
    \textit{Top panels}: $p_\mathrm{min} = 1 \: \frac{\mathrm{GeV}}{c}$.
    \textit{Bottom panels}: $p_\mathrm{min} = 10 \: \frac{\mathrm{GeV}}{c}$.}
    \label{fig:minot_compare}
\end{figure}
In Fig.~\ref{fig:minot_compare} we show the results of our calculation of the $\gamma$-ray source function as a function of photon energy $E_\gamma$.
To calculate the source function we set up power laws in momentum space with varying slope $q$.
Between the panels we vary the minimum momentum $p_\mathrm{min}$ at which the power law starts.
For the top panel this is $p_\mathrm{min} = 1$ GeV c$^{-1}$ and for the bottom panel $p_\mathrm{min} = 10$ GeV c$^{-1}$.
The maximum momentum is fixed at $p_\mathrm{max} = 10^5$ GeV c$^{-1}$.
Each power law is normalized to contain a CRp energy density of $\epsilon_{\mathrm{CR},p} = 1$ erg cm$^{-3}$.
As in the simulation we discretize the spectrum with a total of 12 bins in either case.
To improve convergence we subsample the spectrum by a factor of 10 in momentum space.
We only consider proton-proton collisions for the pion production, thus neglecting the impact of heavier nuclei on the resulting $\gamma$-ray spectrum, and compare our results to the \textsc{minot} package \citep[][]{Adam2020}.
In general, we find reasonable agreement between the two implementations with the symmetric deviation originating from a difference in the low-momentum treatment between \textsc{Crescendo} and \textsc{minot} caused by solving the emission from a power-law in momentum- and energy space, respectively.

\section{Impact of the choice of spectral boundary conditions on the results\label{app:boundary}}
\begin{figure}
    \centering
    \includegraphics[width=\textwidth]{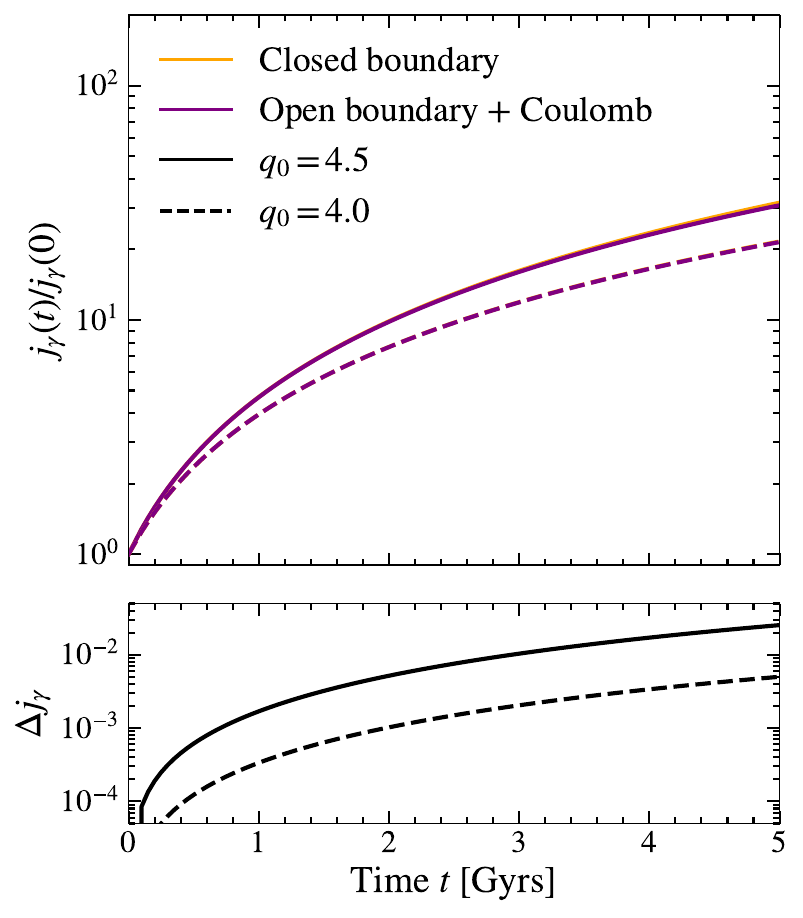}
    \caption{$\gamma$-ray emissivity as a function of time for a spectrum of CRps accelerated at a shock in the periphery of a galaxy cluster, being advected to the center of the galaxy cluster. \textit{Top}: Relative emissivity increase of the spectra when employing a closed lower momentum boundary (red lines), compared to an open lower momentum boundary and Coulomb losses (purple lines). Solid and dashed lines indicate different initial spectral slopes. \textit{Bottom}: Relative error between the runs employing different boundary conditions.}
    \label{fig:boundary}
\end{figure}
To test the impact our choice of a closed lower boundary in the momentum spectrum has on our results, and to compare it to an explicit inclusion of Coulomb losses for protons, we set up a simple toy model.
In this model we initialized proton spectra with initial slopes $q_0 = 4.5$, and $q_0=4$, corresponding to $M_s = 3$, and $M_s \approx 20$, respectively.
We assume that the protons have been accelerated at a shock propagating through a medium with density $n_e = 10^{-4} \: \mathrm{cm}^{-3}$, i.e., the periphery of a galaxy cluster, and are subsequently advected to the center of the galaxy cluster with a density of $n_e = 10^{-3} \: \mathrm{cm}^{-3}$ over the course of 5 Gyr.
The result of this test is shown in Fig.~\ref{fig:boundary}.
We plot the relative change of the $\gamma$-ray emissivity of the spectrum as a function of time in the top panels.
The red line indicates the results when using a closed lower boundary, while the purple line indicates the results when using an open boundary and including Coulomb losses.
Solid lines indicate the runs with $q_0 = 4.5$, while dashed lines indicate the runs with $q_0 = 4.0$.
We find that the difference is minimal, and therefore also plot the relative error $\Delta q_\gamma$ between the two approaches in the bottom panel.
As indicated, the relative error for the spectrum with $q_0 = 4.5$ is of the order 2.5\%, while in the case of $q_0 = 4.0$ it is only of the order 0.5\%.
We attribute this to the very long cooling times of $t_\mathrm{cool} \sim t_\mathrm{Hubble}$ at the pion production threshold energy $T_p \sim 1.22$ GeV.
Hence, for weak shocks, we over-estimate the emission by a few percent, which we accept in light of significantly larger uncertainties in the acceleration efficiency.

%%%%%%%%%%%%%%%%%%%%%%%%%%%%%%%%%%%%%%%%%%%%%%%%%%

\end{document}